\documentclass[twocolumn, secnumarabic, amssymb, aps, showpacs, prd]{revtex4-2}

\usepackage{graphics}
\usepackage[dvips]{graphicx}
\usepackage{dcolumn}
\usepackage{bm}
\usepackage{xcolor}
\usepackage{microtype}
\usepackage{amsmath}
\usepackage{setspace}
\usepackage{mathrsfs}
\begin{document}

\title{Unified Integer and Fractional Quantum Hall Effects from Boundary-Induced Edge-State Quantization}
\author{Pedro Pereyra}
\affiliation{Departamento de Ciencias B\'asicas, UAM-Azcapotzalco, M\'{e}xico D.F., C. P. 02200, M\'exico}
\date{\today}

\begin{abstract}

Despite the success of Landau-level theory and edge-state transport formalisms, a direct microscopic link between bulk quantization and the experimentally observed hierarchy of quantum Hall plateaus has not been established. In particular, no unified microscopic mechanism accounting simultaneously for integer and fractional sequences has been derived within standard quantum mechanics.

Here we show that boundary-induced quantization of edge states provides this missing bridge. Starting from the Landau problem in laterally confined two-dimensional electron systems, we demonstrate that the imposition of Dirichlet, Neumann, and mixed (Robin) boundary conditions discretizes both the guiding-center coordinate and the longitudinal momentum of chiral edge states. The resulting boundary-dependent spectra generate families of edge channels with well-defined multiplicities that couple directly to electronic transport.

When incorporated into an edge-state transport description, this boundary quantization reproduces the integer Hall sequence and simultaneously yields a structured hierarchy of fractional filling factors without invoking separate microscopic mechanisms. We further show that a weak Hall-induced parity-breaking contribution reorganizes the low-energy edge spectrum while leaving the bulk Landau levels intact. This controlled symmetry breaking enhances edge-state multiplicities at small Landau indices and stabilizes the fractional plateaus observed at strong magnetic fields.

The quantized Hall response thus emerges from the interplay between Landau quantization and boundary-induced guiding-center discretization, which together determine the spectrum and occupation of chiral edge channels. These results establish boundary-induced quantization as the microscopic origin of quantum Hall transport and provide a unified description of both integer and fractional regimes within conventional quantum mechanics.
\end{abstract}

\maketitle

\section{Introduction}

The coexistence of integer and fractional quantum Hall plateaus in the same device remains one of the central unresolved microscopic questions in condensed-matter physics. Integer quantization is traditionally attributed to Landau levels of noninteracting electrons, whereas fractional plateaus are explained through strongly correlated many-body states. Although each framework successfully reproduces specific experimental features, their conceptual separation leaves open a fundamental issue: what physical mechanism in real, finite systems generates the universal hierarchy of Hall plateaus and connects bulk quantization to transport?

Since its discovery~\cite{vonKlitzing1980}, the quantum Hall effect has become a paradigmatic example of macroscopic quantization~\cite{Thouless1982,Stormer,Laughlin1983,McIver2020,Yu2010,Hastings2015}. Topological invariants, bulk–edge correspondence, and many-body approaches have clarified the robustness and universality of Hall transport. Yet these descriptions often rely on periodic geometries or treat edge states phenomenologically, leaving the microscopic role of confinement and boundary conditions largely implicit. Consequently, the mechanism by which finite boundaries shape the edge spectrum and produce the observed plateau hierarchy has remained incomplete. The present work does not seek to replace bulk topological or many-body descriptions of the quantum Hall effect; rather, it identifies the role of physically consistent boundary conditions in finite systems as a missing microscopic ingredient linking Landau quantization to the observed plateau hierarchy.

In any physical realization of the quantum Hall effect, the two-dimensional electron system is spatially confined by geometry, electrostatics, and edge structure. Boundaries therefore constitute intrinsic constraints on the electronic wave functions. When a strong magnetic field drives carriers toward the sample edges, Landau states must satisfy boundary conditions that discretize the guiding-center coordinate and the longitudinal momentum of chiral modes. This boundary-induced quantization directly determines the multiplicity and structure of edge channels and therefore the transport spectrum. The absence of a microscopic treatment of this mechanism constitutes a missing bridge between Landau quantization and the hierarchy of Hall plateaus.

Here we show that boundary-induced quantization of edge states provides this bridge. Starting from the Landau problem in laterally confined geometries, we demonstrate that Dirichlet, Neumann, and mixed boundary conditions generate families of chiral edge modes with boundary-dependent multiplicities. Incorporated into an edge-transport description, these spectra reproduce the integer Hall sequence and yield a structured hierarchy of fractional filling factors within the same microscopic description.

We further show that a weak Hall-induced parity-breaking contribution reorganizes the low-energy edge spectrum while preserving the Landau-level ordering that governs the bulk states. The resulting energy shifts are localized near the boundaries, grow with guiding-center displacement, and scale inversely with the Landau index $n$, so their impact is strongest for the lowest levels at strong magnetic fields. In this regime the asymmetry selectively redistributes and pins additional chiral edge branches near the Fermi energy, enhancing the effective edge-state multiplicity and stabilizing the prominent fractional plateaus observed experimentally. Within this description, the integer and fractional quantum Hall effects emerge not as fundamentally distinct phenomena but as complementary manifestations of a single boundary-quantized edge structure shaped by confinement and controlled symmetry breaking.

The aim of this work is to establish a unified analytical description of edge-state quantization in laterally confined two-dimensional electron systems subjected to strong perpendicular magnetic fields. The analysis proceeds at two interconnected levels.

First, we determine how physically consistent boundary conditions imposed on Landau wave functions -- Dirichlet, Neumann, and mixed (Robin-type) conditions -- quantize both guiding-center positions and longitudinal edge momenta. Although these conditions differ in mathematical form and physical interpretation, they produce closely related edge localization and channel densities; their essential differences emerge in the allowed degeneracies and in the resulting filling-factor sequences.

Second, we introduce controlled symmetry breaking through a physically defined parameter associated with Hall-bias--induced edge asymmetry, incorporating structural imbalance, tunneling, and unequal edge transmission. This parameter is not phenomenological: its magnitude is determined by the boundary conditions and transport geometry rather than by fitting. It reorganizes the boundary-quantized edge spectrum while leaving the bulk cyclotron quantization as the underlying structure of the problem. The effect is essential for reproducing the high-field fractional hierarchy. By disentangling boundary-induced multiplicities from genuine parity-breaking effects, the analysis gains both conceptual transparency and predictive power.

In contrast to bulk topological formulations based on periodic boundary conditions or Chern invariants~\cite{Thouless1982,Hastings2015}, the present approach remains entirely within standard quantum mechanics and treats systems with explicit physical boundaries. The resulting boundary-quantized edge spectra feed directly into the Landauer--B\"uttiker transport formalism and lead to a closed analytical expression for the Hall resistance,
\begin{equation}
\rho_{xy}=\frac{h}{e^2\nu_{\rm eff}}, \qquad
\nu_{\rm eff}=\frac{\nu_p}{\nu_c}\nu_n,
\end{equation}
from which both integer and fractional Hall structures follow within a single microscopic description.

\begin{figure*}[hbt]
\begin{center}
\includegraphics[width=17cm]{ 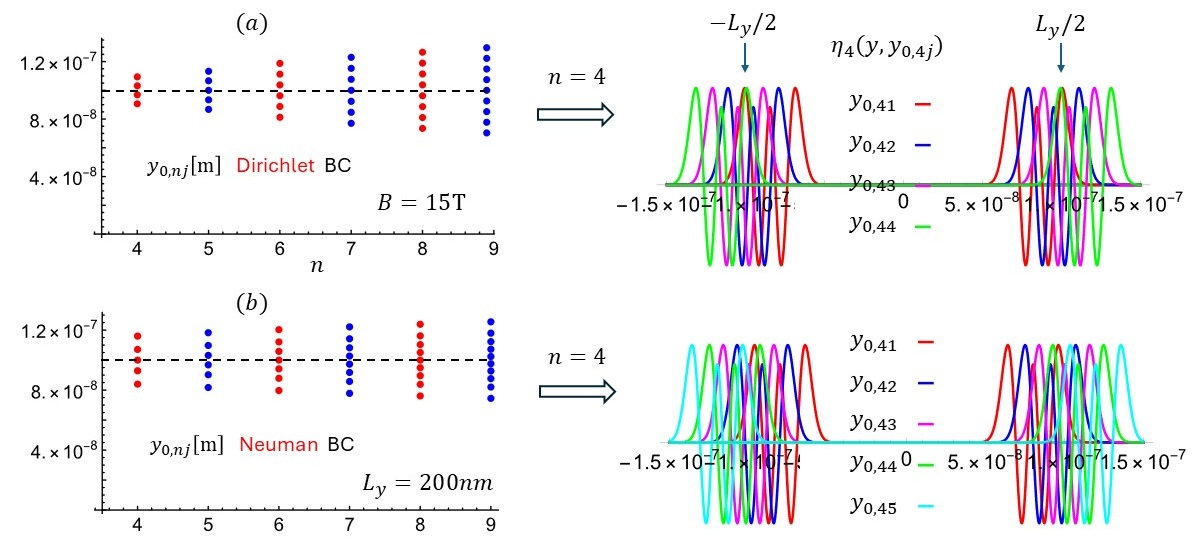}
\caption{Quantized guiding centers obtained from Dirichlet and Neumann boundary conditions, together with the corresponding Landau wave functions evaluated at those centers.
Panels (a) and (b) show the guiding-center positions $y_{0,nj}$ obtained from Eqs.~(\ref{BC_Dirichlet}) and (\ref{BC_Neumann}), respectively.
In the right-hand column, the Landau functions for $n=4$ are plotted at the guiding-center positions indicated in the left column.
The boundary conditions place the guiding centers near the sample edges and produce similar edge localization.
The Robin spectrum, shown in Fig.~\ref{Robinyon}, consists of the Dirichlet spectrum supplemented by two additional guiding-center positions located near the inflection points of the Landau wave packet envelope.
The magnetic field is $B=15$~T and the stripe width is $L_y=0.2~\mu$m.}
\label{yonQuant}
\end{center}
\end{figure*}

Figure~\ref{yonQuant} illustrates the central mechanism: boundary conditions discretize guiding centers near the sample edges and generate families of localized chiral states whose multiplicities depend on the boundary class. These spectra constitute the microscopic input for transport and determine the hierarchy of Hall plateaus.

The remainder of the paper develops this analysis by deriving the boundary-induced quantization of guiding centers and longitudinal momenta, introducing controlled parity breaking, and establishing the connection between edge spectra and macroscopic transport within the Landauer--B\"uttiker formalism, with direct comparison to experiment. The Appendices provide additional derivations and wave-function analyses supporting the main results.

\section{Boundary-Induced Quantization of Edge States}

Quantization in confined quantum systems arises from the interplay between
geometric confinement and admissibility conditions imposed on the wave function.
The confining potential defines the spatial domain accessible to the particle,
while the boundary conditions determine which wave functions are physically
allowed within that domain. When an additional constraint restricts the spatial
structure of admissible states, it becomes the origin of discrete spectra.

In the quantum Hall problem, confinement arises at two distinct and
hierarchically ordered levels. The primary confinement is magnetic:
the perpendicular field quantizes cyclotron motion and produces the Landau
eigenstates, establishing the first level of quantization independently of any
sample boundaries. Lateral confinement associated with the finite width of the
device then acts on these pre-existing states. Its role is not to generate new
transverse eigenmodes, but to define the accessible spatial domain and restrict
the physically allowed positions of the Landau wave packets relative to the edges.

To make this distinction explicit, we represent the Hall bar as a stripe of
width $L_y$ through an auxiliary transverse potential

\begin{equation}
V(y)=
\begin{cases}
0, & |y|<L_y/2,\\
V_{0}, & |y|\ge L_y/2,
\end{cases}
\end{equation}
with $V_{0}$ taken large compared to the Landau-level spacing. This potential
is not intended as a microscopic model of edge electrostatics; it serves only
to define the transverse limits $y=\pm L_y/2$ at which admissibility conditions
must be imposed.

Within the effective-mass and independent-particle approximations, the
Schr\"odinger equation in the Landau gauge $\mathbf{A}=-By\hat{x}$ reads
\begin{equation}\label{SchEq2}
\left[\frac{1}{2m^*}
\left(-i\hbar\nabla-e\mathbf{A}\right)^2 + V(y)\right]\Psi(x,y)
=E\Psi(x,y).
\end{equation}

We first recall the solutions in the absence of lateral confinement.
For $V(y)=0$, the problem is separable, $\Psi(x,y)=e^{ik_x x}\eta(y)$, and
Eq.~(\ref{SchEq2}) reduces to a one-dimensional harmonic oscillator centered at
\begin{equation}
y_0=-k_x\ell_B^2,
\end{equation}
where $\ell_B=\sqrt{\hbar/eB}$ is the magnetic length:
\begin{equation}
\left[
-\frac{d^2}{dy^2}
+\frac{1}{\ell_B^4}(y-y_0)^2
\right]\eta(y)
=\frac{2m^*E}{\hbar^2}\eta(y).
\end{equation}

The normalized eigenfunctions are the displaced Hermite--Gaussian Landau states
\begin{equation}
\eta_n(y)=C_n
\exp\!\left[-\frac{(y-y_0)^2}{2\ell_B^2}\right]
H_n\!\left(\frac{y-y_0}{\ell_B}\right),
\end{equation}
with discrete energies
\begin{equation}\label{LandauEnergy}
E_n=\hbar\omega_c\left(n+\frac{1}{2}\right), \qquad
\omega_c=\frac{eB}{m^*}.
\end{equation}

These transverse states are determined entirely by magnetic quantization and
exist independently of lateral confinement. They are not standing waves created
by the confining potential, but localized Landau wave packets whose guiding-center
coordinate $y_0$ can take any value in an extended system. Lateral finiteness
therefore does not generate new transverse modes; it restricts the spatial
locations available to these pre-existing states relative to the sample edges.

When the transverse states already possess a fixed internal structure,
boundary conditions select the configurations compatible with the finite
geometry and thereby act as an additional quantization mechanism.

The present problem belongs to this class. Landau eigenfunctions are
Hermite--Gaussian packets characterized by a discrete set of nodes, extrema,
and inflection points. Imposing boundary conditions therefore does not merely
truncate the wave function; it determines which of these structural features
can coincide with the edges and thus directly selects the allowed guiding-center
positions and the associated longitudinal momenta.

In a stripe geometry, admissibility must be enforced at
$y=\pm L_y/2$. Different boundary prescriptions select different geometric
features of the Hermite--Gaussian envelope:

\begin{itemize}
\item Dirichlet conditions align the zeros of the Hermite polynomial with the
boundaries.
\item Neumann conditions pin the extrema of the wave packet at the edges.
\item Robin conditions constrain its curvature and effectively select outer
inflection structures.
\end{itemize}

Because these structural features occur with comparable spatial density, the
resulting guiding-center locations are similar across boundary classes, while
their multiplicities differ in a systematic and physically transparent manner.
The detailed shape of the confining potential therefore becomes secondary: its
essential role is to define the transverse domain, whereas the boundary
conditions determine the selected spectrum.

We now make these boundary constraints explicit. The transverse Landau
wave functions must satisfy boundary conditions at $y=\pm L_y/2$ of the form
\begin{align}
\text{Dirichlet:}\qquad
\left.\eta_n\right|_{y=\pm L_y/2}&=0, \label{BC_Dirichlet}\\[4pt]
\text{Neumann:}\qquad
\left.\frac{d\eta_n}{dy}\right|_{y=\pm L_y/2}&=0, \label{BC_Neumann}\\[4pt]
\text{Robin:}\qquad
\left.\left(\eta_n+\frac{d\eta_n}{dy}\right)\right|_{y=\pm L_y/2}&=0.
\label{BC_Robin}
\end{align}

For the unit-coefficient choice adopted here, the Robin condition is
equivalent to a curvature constraint; a short derivation is provided in
Appendix A.
The Neumann and Robin conditions follow from the requirement that the
probability current remain tangent to the boundary, so that its normal
component vanishes at the edges,\cite{Morse1953} whereas the Dirichlet condition is imposed
directly as a boundary admissibility constraint.

Imposing any of these conditions at both edges restricts the allowed
positions of the guiding center and therefore quantizes the longitudinal
momentum
$k_x$. The boundary does not generate new transverse modes; instead, it selects
a discrete set of admissible Hermite--Gaussian wave packets localized near the
edges. The resulting channel multiplicities depend on which structural feature
of the wave packet is pinned at the boundaries and thus on the boundary class
itself.

To implement this quantization explicitly, it is convenient to introduce the
dimensionless variable
\begin{equation}\label{defxi}
\xi=\frac{\pm L_y/2-y_0}{\ell_B}.
\end{equation}
The boundary condition determines the roots $\xi_{nj}$ and thus yields a discrete
spectrum of guiding centers for any effective width $L_y$.
Using $k_x=-y_0/\ell_B^2$, the corresponding quantized longitudinal momenta are
\begin{equation}\label{defkxnj}
k_{x,nj}=\frac{1}{\ell_B}
\left(\xi_{nj}\pm\frac{L_y}{2\ell_B}\right).
\end{equation}

As illustrated in Fig.~\ref{yonQuant}, discretized guiding-center positions for $n=4,5,\ldots,9$ are shown together with the localized eigen-wave packets near the edge at $L_y/2$. The right column displays the transverse wave packets $\eta_{4j}(y)$ for $n=4$: the upper panel corresponds to Dirichlet boundary conditions with $j=1,\ldots,4$, while the lower panel corresponds to Neumann boundary conditions with $j=1,\ldots,5$. Panel~(a) shows that Dirichlet boundary conditions yield $n$ guiding centers for the $n$th Landau level, whereas panel~(b) shows that Neumann boundary conditions increase this number to $n+1$; the corresponding Robin spectrum, presented in Appendix~B, contains $n+2$ solutions. Although the resulting spectra are closely related for Dirichlet, Neumann, and Robin conditions, they exhibit systematic differences in edge-channel multiplicity.

These multiplicities translate directly into different effective channel
weights in transport.
When normalized to the current carried by a Dirichlet edge channel,
a Neumann channel contributes a fraction $n/(n+1)$ of that current, while a
Robin channel contributes $n/(n+2)$.
These fractional factors arise purely from boundary-induced channel counting
and anticipate the effective filling factors that emerge within the
Landauer--B\"uttiker description.
Their physical interpretation in terms of boundary-induced quantization and
vanishing normal current is discussed in Appendix~B.

The associated transverse magnetic confinement energy of each edge
channel is
\begin{equation}\label{Undef}
U_{nj}=\frac{\hbar\omega_c}{2}\,\xi_{nj}^2,
\end{equation}
and the total energy can be written as $E_{nj}=K_{nj}+U_{nj}$, where
$K_{nj}=\hbar^2 k_{x,nj}^2/2m^*$.

\begin{figure*}[hbt]
\begin{center}
\includegraphics[width=16.5cm]{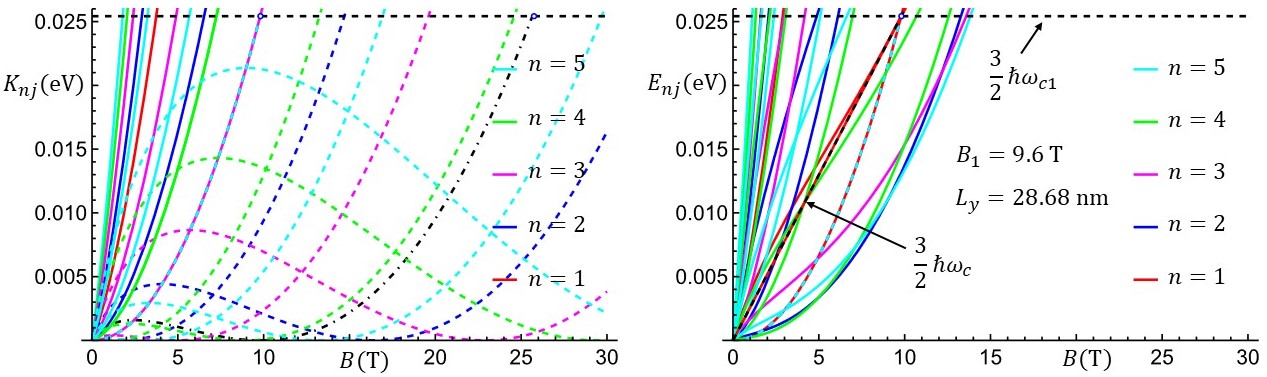}
\caption{Kinetic and total energies of edge channels as functions of magnetic field.
Left panel: longitudinal kinetic energies
$K_{nj}=\hbar^2 k_{x,nj}^2/2m^*$
for the first five Landau levels, computed from the exact quantized momenta in Eq.~(\ref{defkxnj}).
Approximately half of the kinetic branches remain below the Fermi energy.
Right panel: total energies
$E_{nj}=K_{nj}+U_{nj}$.
Including the magnetic confinement potential $U_{nj}$ reorders the spectrum:
branches associated with large $|\xi_{nj}|$ are shifted upward, while those with
small or vanishing $\xi_{nj}$ experience smaller energy shifts.
As a result, fewer channels remain below $E_F$, whereas the $\xi_{nj}=0$ branches
remain purely kinetic and evolve together with the $n=1$ Landau level.
The persistence (pinning) of selected channels after the bulk Landau level crosses
$E_F$ constitutes a microscopic mechanism responsible for residual chiral edge
channels and, consequently, for the emergence of fractional Hall plateaus.
Parameters are indicated in the plots; details on the effective stripe width
$L_y$ are provided in the Supplementary Information.}
\label{kineticenergies}
\end{center}
\end{figure*}

Importantly, this geometric quantization does not modify the Landau energy
$E_n$ in Eq.~(\ref{LandauEnergy}), but redistributes it between longitudinal
kinetic and transverse magnetic confinement contributions within a given
Landau level.

Only those edge channels whose total energy lies sufficiently close to the
Fermi energy contribute to transport, establishing the microscopic link
between boundary-induced quantization and quantized Hall conductance.

To visualize this redistribution and to make explicit the relation between the
Landau index $n$ and the discretized longitudinal momenta $k_{x,nj}$, we now
introduce approximate expressions for the guiding-center eigenvalues $\xi_{nj}$
obtained under the Robin boundary condition.
These approximations provide analytical insight into the channel structure and
are instrumental for the Landauer--B\"uttiker analysis that follows.

For both even and odd $n$, the approximate eigenvalues can be written as
\begin{equation}\label{aproxyoj}
\xi_{nj}\simeq
\mp\sqrt{2n+1}
\pm (j-1)\frac{\Delta y_{0,nj}}{\ell_B},
\qquad
j=1,2,\ldots,n+2,
\end{equation}
where the spacing between adjacent guiding centers is
\begin{equation}\label{aproxDeltaj}
\Delta y_{0,nj} \simeq
\ell_B\,\frac{2\sqrt{2n+1}}{n+1}.
\end{equation}
These expressions converge rapidly to the exact numerical roots for fixed
$j$ as $n$ increases.
The corresponding approximate longitudinal momenta for the Robin condition are
\begin{equation}\label{aproxkxn}
k_{x,nj}
=
-\frac{y_{0,nj}}{\ell_B^2}
\simeq
\pm\,\frac{\sqrt{2n+1}}{\ell_B}
+\frac{(j-1)\,\Delta y_{0,nj}}{\ell_B^2}
\mp\frac{L_y}{2\ell_B^2},
\end{equation}
yielding $n+2$ discrete momentum values for each Landau level.

These analytical expressions allow one to follow directly the evolution of edge
channels with magnetic field.
The interplay between the longitudinal kinetic energy $K_{nj}$ and the transverse
magnetic confinement energy $U_{nj}$ determines which channels remain active in
transport as the field increases.
Details on the relation between momentum sign, edge localization, and channel
exchange are provided in Appendix C.

Figure~\ref{kineticenergies} compares the kinetic energies
$K_{nj}=(\hbar^2/2m^*)k_{x,nj}^2$ with the corresponding total energies
$E_{nj}=K_{nj}+U_{nj}$.
While the kinetic spectrum alone exhibits multiple low-energy branches, the
inclusion of $U_{nj}$ selectively shifts channels with large $|\xi_{nj}|$
upward, leaving only those with small or vanishing $|\xi_{nj}|$ below the Fermi
energy.

The $\xi_{nj}=0$ modes, which occur for odd Landau indices, remain purely kinetic
and evolve together with the $n=1$ Landau level.
These pinned chiral channels preserve current continuity even after the bulk
Landau level crosses $E_F$.
Thus, although magnetic confinement progressively reduces the number of
conducting channels, it does not eliminate them completely.
The surviving edge states provide the microscopic basis for the residual
fractional Hall plateaus observed experimentally, as made explicit in the
transport analysis that follows.

This correspondence between guiding-center quantization, energy redistribution,
and channel survival establishes the microscopic foundation for the Hall
resistance derived within the Landauer--B\"uttiker framework.
While boundary conditions alone account for most observed Hall plateaus, the
next section incorporates the effect of the Hall potential, which further
refines the plateau structure.

\section{Edge Spectra from Boundary Conditions and Parity Breaking}

In the previous section we analyzed the consequences of boundary conditions alone
on the quantization of guiding centers and the resulting multiplicity of edge
states.
Imposing Dirichlet, Neumann, or Robin boundary conditions on Landau wave
functions in a laterally finite system produces families of edge states whose
number scales with the Landau index, with small but systematic differences among
the three cases.
These boundary-induced multiplicities generate a rich hierarchy of edge
channels, providing essential microscopic input required for the quantization of Hall
transport.

We now extend this analysis by incorporating, in addition to boundary conditions,
a controlled parity-breaking mechanism that reflects the fact that, under Hall
bias, the two edges of the sample are not energetically equivalent, even though
both remain sharply confined.
In the quantum Hall regime, chiral edge states and the strong suppression of
backscattering are essential ingredients in the emergence of quantized Hall
resistance.
Within the Landauer--B\"uttiker framework, transport is described in terms of
edge channels propagating along the upper and lower boundaries, characterized by
chemical potentials $\mu_A$ and $\mu_B$.
Their difference defines the Hall potential $V_H=(\mu_A-\mu_B)/e$.

In idealized treatments of open quantum Hall systems, the Hall potential is
introduced at the level of electrochemical transport and does not explicitly
enter the single-particle Schr\"odinger equation.
Here we retain a small, controlled fraction of the associated transverse force
in order to describe the microscopic asymmetry between the two edges.
This residual contribution captures the fact that, under Hall bias, the edges
are not energetically equivalent even though both remain sharply confined.

Specifically, we introduce a weak linear term in the transverse potential,
proportional to the coordinate $y$,
\begin{equation}
V(y)=
\begin{cases}
b\, e B v_x\, y, & |y|<L_y/2,\\
V_{0}, & |y|\ge L_y/2,
\end{cases}
\end{equation}
where $0<b\ll1$ is a dimensionless parameter quantifying the fraction of the
Hall force retained in the transverse dynamics.
This term tilts the magnetic confinement potential and breaks the parity
symmetry $y\rightarrow -y$ of the ideal Landau problem.
Using $v_x=\hbar k_x/m^*=-y_0\omega_c$, the parity-breaking contribution can be
expressed entirely in terms of the guiding-center coordinate $y_0$.

With this modification, the transverse Schr\"odinger equation becomes
\begin{equation}\label{ParitySchEq}
\left[
-\frac{d^2}{dy^2}
+\frac{1}{\ell_B^4}(y-y_0)^2
+ b\,\frac{y_0}{\ell_B^4}\,y
\right]\eta(y)=\frac{2m^*E}{\hbar^2}\eta(y),
\end{equation}
where $y_0=-k_x\ell_B^2$ is the guiding-center coordinate.
The equation remains exactly solvable and preserves the Landau-level indexing,
while shifting the effective oscillator center and redistributing the energies
of edge-localized modes within each Landau level.

The normalized solutions can be written in the compact form
\begin{equation},\label{PBsolution}
\eta(y)=
c_n\,
\exp\!\left[
-\frac{(y-(1+b)y_0)^2}{2\ell_B^2}
\right]
H_{n'}\!\left(\frac{y-(1+b)y_0}{\ell_B}\right),
\end{equation}
where the effective polynomial index is
\begin{equation}\label{Landaunp}
n'(n,b)=n+ \frac{b(b+2)}{2n}\,\frac{L_y}{\ell_{B_1}}.
\end{equation}

The quantity $n'$ is not a new Landau quantum number but an effective index
characterizing boundary-compatible solutions of the parity-broken problem.
Integer values of $n'$ identify the transverse wave packets that satisfy the
boundary conditions and therefore correspond to physically realizable edge
eigenstates. Here $\ell_{B_1}$ denotes the magnetic length evaluated at the magnetic field
corresponding to the $n=1$ Hall plateau.
The corresponding energy eigenvalues are
\begin{equation}\label{PBenergy}
E=\hbar\omega_c\!\left(n+\frac{1}{2}\right)
- b(b+2)\,\frac{y_0^2}{\ell_B^2}\,\hbar\omega_c .
\end{equation}

Several immediate consequences follow.
The magnitude of the parity-breaking parameter required to reproduce the
fractional sequences discussed below is small, so that the Landau-level ordering
remains the organizing structure of the spectrum.
The associated energy shift grows with guiding-center displacement
(equivalently with $k_x^2\ell_B^2$), localizing its impact near the sample
boundaries and making it increasingly relevant at strong magnetic fields.
In addition, the shift scales inversely with the Landau index $n$, so that its
spectral effect is strongest for the lowest levels.
As a result, parity breaking does not generate new edge modes but selectively
stabilizes additional low-energy branches already permitted by boundary
quantization, thereby enhancing the effective edge-state multiplicity.
\begin{figure}[hbt]
\begin{center}
\includegraphics[width=8.5cm]{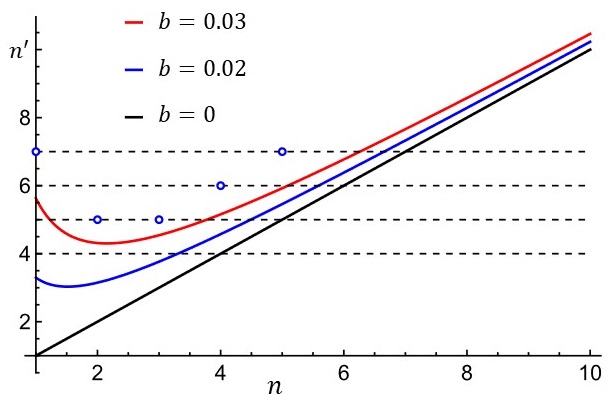}
\caption{
Effective eigenfunction index $n'$ and corresponding number of edge states,
obtained under the Neumann boundary condition, as functions of the Landau index
$n$ and the parity-breaking parameter $b$.
The solid curves show the analytical expression of
$
n'(n,b)
$ in (\ref{Landaunp}).
Horizontal dashed lines indicate integer values of $n'$.
Small circles denote the number of zeros of the Neumann boundary condition,
$\left.d\eta/dy\right|_{y=\pm L_y/2}=0$, which correspond to local extrema of the
Hermite--Gaussian envelope where the guiding centers are pinned by the boundary.
For finite parity breaking (e.g.\ $b\simeq0.03$), the resulting integer sequences
exhibit a non-monotonic dependence on $n$, signaling an enhancement of low-energy
edge-state multiplicity at small Landau indices.}
\label{parityFig}
\end{center}
\end{figure}

A convenient way to quantify this effect is through the effective Hamiltonian
order $n'$, whose integer part determines the number of available low-energy edge
states for a given Landau index and parity-breaking strength.
Equation~(\ref{Landaunp}) shows explicitly that parity breaking enhances the
number of edge channels most strongly for $n=1$ and $n=2$, and that this
enhancement depends nontrivially on $b$.

This behavior is illustrated in Fig.~\ref{parityFig}, which shows $n'(n,b)$ as a
function of the Landau index for representative values of $b$.
For intermediate parity breaking, $n'(n,b)$ exhibits a non-monotonic dependence
on $n$.
The intersections with integer values of $n'$, indicated by discrete points,
correspond to the actual number of edge branches contributing at the transport
energy scale.
These sequences anticipate the channel multiplicities that enter the effective
filling factors and the fractional Hall plateaus derived below.

At this stage, the role of parity breaking is purely microscopic.
It reorganizes the edge spectrum by stabilizing additional long-lived
low-energy branches without modifying the bulk Landau quantization.
Figure~\ref{parityFig} shows, for example, the sequence $(7,5,5,6,7,\ldots)$ of
zeros of the Neumann boundary condition, corresponding to local extrema of the
Landau wave function at the edge.
Through the effective filling factor $\nu_{\rm eff}$, this sequence generates
nearly the complete set of prominent fractional Hall plateaus observed
experimentally in high-mobility samples.

Within this framework, the fractional quantum Hall effect emerges as the
cumulative consequence of two single-particle mechanisms:
(i) boundary-induced discretization of guiding centers and longitudinal momenta,
and (ii) weak parity breaking that selectively enhances low-energy edge-state
multiplicities.
In the next section, these modified edge spectra are translated directly into
quantized Hall resistance plateaus within the Landauer--B\"uttiker transport
formalism.

\begin{figure*}[hbt]
\begin{center}
\includegraphics[width=10.8cm]{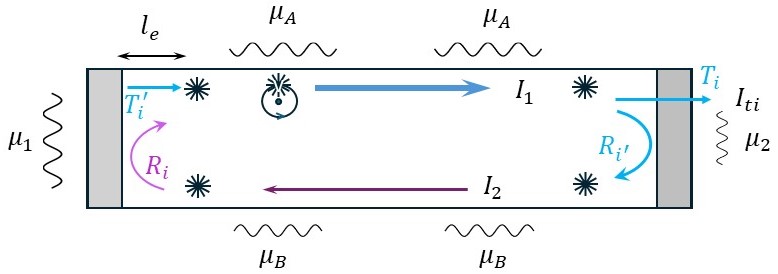}
\caption{
Schematic representation of the Landauer–Büttiker transport picture applied to
the parity-quantized Landau spectrum.
Electrons are injected from the left reservoir (chemical potential $\mu_1$) into the quantized edge channels whose total energies $E_{nj}$ match the Fermi level $E_F$; resonant tunneling across the contact barrier selectively populates these discrete states, while electrons injected into non-resonant states relax inelastically toward them.
Inside the sample, edge electrons propagate chirally along equipotential contours with constant longitudinal momentum $k_{x,nj}$ and strongly suppressed backscattering.
At the drain contact (chemical potential $\mu_2$), part of the incoming flux is transmitted into the reservoir (forming the transmitted current $I_t$) and part is reflected back into lower edge channels, contributing to the reflected edge flux.
The transmission coefficient $T_j$ for individual channel $j$ results from the sum over reservoir feeding states, and the net transmission summed over the $\nu_p$ populated channels equals the $T_{1p}$ used in the text (see Eqs.~\ref{T1}--\ref{FC}), providing the flux-conservation link between microscopic channel transmissions and the macroscopic currents.
 } \label{QHallEffect}
\end{center}
\end{figure*}

\section{Edge Channel Transport and Hall Quantization}

Having established that physically consistent boundary conditions and controlled
parity breaking generate a discrete hierarchy of chiral edge channels, we now
examine how this microscopic edge structure translates into measurable
transport.

Before introducing a specific transport formalism, it is essential to clarify
how the counting, occupation, and energetic ordering of boundary-quantized edge
states evolve as the magnetic field and Fermi energy are varied. In a finite Hall
geometry, electrical conduction is mediated by boundary-confined states whose
quantized longitudinal momenta determine both the number of available channels
and their energetic alignment with the Fermi level.

\subsection{Boundary- and Symmetry-Induced Filling-Factor Hierarchies}

At this stage the focus is not yet on transport coefficients but on the
microscopic organization of the edge spectrum itself. The relevant quantity is
the edge filling configuration, which reflects the partial occupation of
boundary-quantized chiral channels associated with successive Landau levels and
their edge-mode families. Boundary conditions fix the allowed guiding-center
positions and therefore the channel multiplicities, while parity breaking
selectively lifts degeneracies at low Landau indices. Together, these mechanisms
generate a structured hierarchy of edge configurations that anticipates both
integer and fractional Hall plateaus.

Substituting the quantized longitudinal momenta into the total energy expression
yields the discrete edge-state energies
\begin{eqnarray}\label{EnergyEn}
E_{nj} &=&
\frac{\hbar^2}{2m^*}\frac{1}{\ell_{B}^2}
\left(\xi_{nj} \pm \frac{L_y}{2\ell_{B}}\right)^2
+ \frac{\hbar \omega_c}{2}\xi_{nj}^2 .
\end{eqnarray}

For the present discussion, the explicit analytical form of $E_{nj}$ is less
important than its ordering and spacing. The discrete set of energies defines the
hierarchy of edge channels associated with each Landau level and determines which
of them lie closest to the Fermi energy as the magnetic field varies.

The separation between neighboring edge branches introduces a characteristic
energy scale, typically of order $\hbar\omega_c/\sqrt{n+1}$, which governs the
progressive activation and depopulation of channels. Low-energy branches remain
pinned near the Fermi level over finite magnetic-field intervals, while higher
branches shift away more rapidly. This energetic hierarchy provides the
microscopic basis for the emergence of plateau structures.

We first distinguish the number of bulk Landau-level families that are
energetically available to supply edge states. This quantity,
\[
\nu_n \equiv n_{\rm max},
\]
counts the highest Landau index intersecting the Fermi energy and represents the
reservoir of edge-mode families generated by magnetic quantization.
It is a bulk spectral quantity and does not coincide with the experimentally
measured Hall filling factor.

Boundary conditions then determine how many edge channels are realized from each
Landau family. The total number of available chiral channels is therefore
\[
\nu_c =
\begin{cases}
n_{\rm max}, & \text{Dirichlet boundary condition},\\[4pt]
n_{\rm max} + 1, & \text{Neumann boundary condition},\\[4pt]
n_{\rm max} + 2, & \text{Robin boundary condition}.
\end{cases}
\]

Parity breaking further redistributes the energies of these channels and enhances
their multiplicity at small Landau indices, without modifying the underlying
Landau-level structure. The combined action of boundary constraints and symmetry
breaking therefore produces a discrete hierarchy of admissible edge-state
configurations.

Within this hierarchy, the experimentally relevant filling factors arise from the
energetic alignment of boundary-quantized channels with the Fermi level. The
ordering and spacing of the edge-state energies determine which channels remain
active as the magnetic field varies.

An immediate consequence is that the spacing between effective filling
configurations is not imposed phenomenologically but follows from the density of
boundary-allowed channel configurations. Plateau widths therefore follow
directly from the discrete structure of the guiding-center spectrum and its
multiplicities: densely spaced configurations produce narrow plateaus, while
sparser regions generate broader ones.

This boundary- and symmetry-induced hierarchy provides the microscopic input
required for transport. The translation from channel occupation to measurable
conductance is carried out within the Landauer--B\"uttiker framework, which we
now introduce.

\subsection{Landauer--B\"uttiker formulation with parity-dependent edge channels}

Before introducing the explicit Landauer--B\"uttiker transport relations, it is
useful to clarify the physical mechanism by which boundary-quantized edge states
become populated and contribute to current flow.

In the Landauer--B\"uttiker picture, electrons in the source reservoir occupy an
almost continuous distribution of states around the Fermi energy $E_F$, thermally
broadened by $\sim k_B T$. When the total energy of an edge state,
$E_{nj}=K_{nj}+U_{nj}$, aligns with $E_F$, electrons within this thermal window
can tunnel across the contact barrier and, after inelastic relaxation, populate
the quantized edge state $j$ with an effective probability
$T'_j=\sum_{i=1}^{M}T'_{ji}$ determined by the contact transmission matrix
elements rather than by bulk occupation statistics.

Because the edge states belonging to a given Landau level have slightly
different energies, their population evolves systematically as the magnetic
field varies. Lower-energy branches become occupied later and remain localized
near the Fermi level, while higher-energy branches depopulate and shift upward
in energy. Once populated, these edge channels behave as effectively
one-dimensional, phase-coherent chiral conduits linking source and drain.

Electrons propagate along equipotential contours with nearly constant
longitudinal momentum $k_{x,nj}$ and strongly suppressed backscattering. At the
drain contact ($\mu_2$), a fraction of the incident flux is transmitted into the
reservoir, contributing to the total current $I_t$, while the remainder is
reflected into lower-energy edge channels along the opposite boundary. This
partial transmission and reflection process is quantitatively described by the
contact transmission and reflection coefficients.

The natural current scale in this regime is set by the number of bulk
Landau-level families available to supply edge conduction. Using the
Landau-index filling introduced above, $\nu_n = n_{\rm max}$, we identify the
number of occupied Landau levels capable of feeding edge-mode families.

Each such Landau family provides, under ideal injection, the capacity to carry
one quantum of conductance. In the phase-coherent limit, the ideal edge current
$I^{\rm ed}$ injected along the upper boundary at chemical potential $\mu_A$,
referenced to $\mu_2$, is therefore
\[
I^{\rm ed} = \frac{e}{h}\,\nu_n\,(\mu_A-\mu_2).
\]
This quantity represents the intrinsic injection scale set by magnetic
quantization; the measurable transmitted current is subsequently determined by
the boundary-resolved channel multiplicities and the corresponding transmission
and reflection processes at the contacts.

Once the boundary-quantized edge channels and their energetic ordering are
established, their contribution to electrical transport can be treated in a
direct and unambiguous manner within the Landauer--B\"uttiker formalism.
This framework is particularly appropriate here because current is carried by
chiral edge modes whose number, multiplicity, and alignment with the Fermi level
are fixed microscopically by boundary conditions and controlled parity breaking.

Unlike bulk linear-response approaches, the Landauer--B\"uttiker picture connects
discrete edge spectra to measurable currents through transmission and reflection
processes at the contacts, without requiring disorder-induced localization or
spatial averaging.

Following Ref.~\onlinecite{Buttiker1988}, the transmission and reflection
coefficients between the reservoirs and the populated edge channels at contact 1 are

\begin{equation}\label{T1}
T_{1p}=\sum_{j=1}^{M}\sum_{i=1}^{\nu_p}T_{ji}, \qquad
R_{1p}=\sum_{i=1}^{\nu_p}\sum_{l=1}^{\nu_p}R_{i'l},
\end{equation}
where $\nu_p$ denotes the number of edge channels populated at the Fermi level.
Similarly, at contact 2,
\begin{equation}\label{T2}
T_{2p}=\sum_{j=1}^{M'}\sum_{l=1}^{\nu_p}T_{jl}, \qquad
R_{2p}=\sum_{l=1}^{\nu_p}\sum_{l'=1}^{\nu_p}R_{l'l}.
\end{equation}
Flux conservation requires
\begin{equation}\label{FC}
T_{1p}+R_{1p}=\nu_p, \qquad T_{2p}+R_{2p}=\nu_p.
\end{equation}

Within the Landauer description, the current injected from a reservoir at
chemical potential $\mu$ into the $j$th populated edge channel is
\begin{equation}\label{LandauerCurr}
I_j=\frac{e}{h}(\mu-\mu_0)T_j ,
\end{equation}
where $T_j$ is the transmission probability into that channel and
$\partial n_e/\partial E_j=1/hv_j$ has been used for the one-dimensional
density of states.

The total current leaving contact 1 contains both the injected component and the
portion reflected from the opposite boundary,
\begin{equation}\label{CurrI1a}
I_{1p}=\frac{e}{h}T_{1p}(\mu_1-\mu_2)+I^{\mathrm{ed}}_{2R}.
\end{equation}

The edge current carried by the populated channels can be expressed in terms of
the ideal edge-current scale $I^{\rm ed}$ introduced above. Because only a
subset $\nu_p$ of the $\nu_c$ boundary-allowed channels is energetically aligned
with the Fermi level, the injected edge current is reduced by the ratio
$\nu_p/\nu_c$. One obtains
\begin{equation}\label{CurrI1b}
I_{1p}^{\mathrm{ed}}
=\frac{\nu_p}{\nu_c}I^{\rm ed}
=\frac{e}{h}\frac{\nu_p}{\nu_c}\nu_n(\mu_A-\mu_2),
\end{equation}
To simplify the notation and to isolate the physically relevant combination of
microscopic ingredients, we introduce the effective filling factor
\begin{equation}\label{fillingnu}
\nu_{\rm eff}=\frac{\nu_p}{\nu_c}\,\nu_n ,
\end{equation}
which combines three distinct contributions: the bulk Landau-level supply
($\nu_n$), the boundary-induced channel multiplicity ($\nu_c$), and the number of
channels actually populated near the Fermi level ($\nu_p$).

Using the flux-conservation relations~(\ref{FC}), Eq.~(\ref{CurrI1b}) can be
rewritten as
\begin{eqnarray}\label{CurrI1b2}
I_{1p}^{\mathrm{ed}}
&=&\frac{e}{h}\frac{\nu_{\rm eff}}{\nu_p}(T_{2p}+R_{2p})(\mu_A-\mu_2) \cr
&=& I_{1t}+I_{1R},
\end{eqnarray}
with transmitted and reflected contributions
\begin{equation}\label{currI1t}
I_{1t}=\frac{e}{h}\frac{\nu_{\rm eff}}{\nu_p}T_{2p}(\mu_A-\mu_2),
\end{equation}
and
\begin{equation}\label{currI1R}
I_{1R}=\frac{e}{h}\frac{\nu_{\rm eff}}{\nu_p}R_{2p}(\mu_A-\mu_2).
\end{equation}

At contact 2, the edge current reads
\begin{eqnarray}\label{CurrI2a}
I_{2p}^{\mathrm{ed}}
&=&\frac{e}{h}\nu_{\rm eff}(\mu_B-\mu_2) \cr
&=&\frac{e}{h}\frac{\nu_{\rm eff}}{\nu_p}(T_{1p}+R_{1p})(\mu_B-\mu_2).
\end{eqnarray}

Combining these relations yields
\begin{equation}\label{Comb1}
T_{1p}(\mu_1-\mu_2)
+\frac{\nu_{\rm eff}}{\nu_p}R_{1p}(\mu_B-\mu_2)
=\nu_{\rm eff}(\mu_A-\mu_2),
\end{equation}
and
\begin{equation}\label{Comb2}
(\mu_B-\mu_2)=\frac{R_{2p}}{\nu_p}(\mu_A-\mu_2).
\end{equation}

From these expressions,
\begin{equation}\label{muA2}
\mu_A-\mu_2=
\frac{\nu_p^2}{\nu_{\rm eff}}
\frac{T_{1p}}{\nu_p^2-R_{1p}R_{2p}}
(\mu_1-\mu_2),
\end{equation}
and
\begin{equation}\label{muB2}
\mu_B-\mu_2=
\frac{\nu_p}{\nu_{\rm eff}}
\frac{T_{1p}R_{2p}}{\nu_p^2-R_{1p}R_{2p}}
(\mu_1-\mu_2).
\end{equation}

\subsection{The Quantum Hall Resistance}

With the microscopic edge spectrum fixed by boundary conditions and parity
breaking, the Hall resistance follows directly from the imbalance of chemical
potentials established between counterpropagating edge channels.

From Eqs.~(\ref{muA2}) and (\ref{muB2}) we obtain
\begin{equation}\label{HallVoltageFinal}
\mu_A-\mu_B=
\frac{1}{\nu_{\rm eff}}
\frac{T_{1p}\nu_p(\nu_p-R_{2p})}{\nu_p^2-R_{1p}R_{2p}}
(\mu_1-\mu_2),
\end{equation}
and the transmitted current
\begin{equation}\label{CurrFinal}
I=
\frac{e}{h}
\frac{\nu_pT_{1p}T_{2p}}{\nu_p^2-R_{1p}R_{2p}}
(\mu_1-\mu_2).
\end{equation}

The Hall voltage $V_H=(\mu_A-\mu_B)/e$ and the current $I$ therefore satisfy
\begin{equation}\label{VH}
V_H=
\frac{1}{e\nu_{\rm eff}}
\frac{T_{1p}\nu_pT_{2p}}{\nu_p^2-R_{1p}R_{2p}}
(\mu_1-\mu_2),
\end{equation}
and
\begin{equation}\label{Itrans}
I=
\frac{e}{h}
\frac{\nu_pT_{1p}T_{2p}}{\nu_p^2-R_{1p}R_{2p}}
(\mu_1-\mu_2).
\end{equation}

Eliminating the contact-dependent transmission and reflection coefficients
from Eqs.~(\ref{VH}) and (\ref{Itrans}) yields the universal relation
\begin{equation}\label{VHrho}
V_H=\frac{h}{e^2}\frac{I}{\nu_{\rm eff}},
\end{equation}
which is independent of microscopic contact details and depends only on the
effective filling factor defined by the boundary-quantized edge spectrum.
The Hall resistivity therefore takes the compact form
\begin{equation}\label{rhoXY}
\rho_{xy}=\frac{h}{e^2}\frac{\nu_c}{\nu_p\nu_n}.
\end{equation}

This result expresses the quantized Hall resistance directly in terms of the
microscopic hierarchy established earlier. Three integers determine the
quantization pattern: the number of Landau-level families supplying edge modes
($\nu_n$), the number of boundary-allowed channels set by confinement
($\nu_c$), and the number of channels effectively populated near the Fermi level
($\nu_p$). All three follow from magnetic quantization, boundary constraints, and
controlled parity breaking, without introducing phenomenological transport
parameters.

Because $\nu_c$ and $\nu_p$ arise from boundary-induced multiplicities and
parity-controlled channel selection, their ratios naturally generate both
integer and fractional values of $\nu_{\rm eff}$ within the same microscopic
framework.

\begin{figure*}[hbt]
\begin{center}
\includegraphics[width=14.5cm]{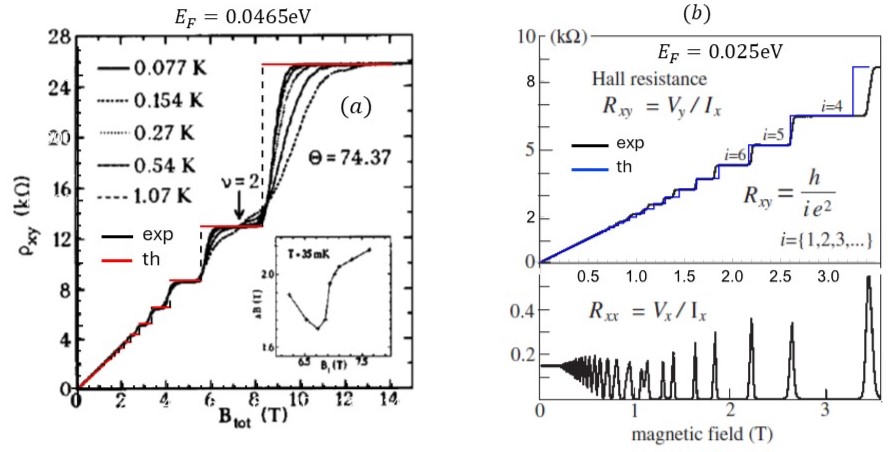}
\caption{Experimental and theoretical Hall resistance (in units of $h/e^2$).
(a) Theoretical prediction compared with the experimental Hall-resistance curve
measured in a two-dimensional electron system at a GaInAs/InP interface, reported
in Ref.~\cite{Koch1993} (reproduced with permission from Elsevier).
(b) Comparison between theory and experiment for a two-dimensional electron system
in a GaAs/(AlGa)As heterostructure, reported in Ref.~\cite{WeisVonKlitzing}
(reproduced with permission from The Royal Society, UK).
Theoretical curves were obtained from Eq.~(\ref{Hallreslfloor}) using the boundary-
determined effective filling factor for the Fermi energies indicated in each
panel, without additional fitting parameters.
Both integer and fractional plateaus arise from the same boundary-quantized edge
spectrum.}
\label{graphsIQHE}
\end{center}
\end{figure*}

This completes the microscopic–to–transport connection: boundary conditions alone determine the admissible edge-channel multiplicities and yield a first-principles prediction of the Hall resistance, while controlled parity breaking refines the spectrum and produces the corrected effective filling factor $\nu_{\rm eff}$, which, as shown in the next section, quantitatively reproduces the observed plateau hierarchy.

\section{Integer and Fractional Hall Plateaus from Boundary-Quantized Edge Transport}

We now determine explicitly the filling-factor sequences generated by each
class of boundary condition.

\paragraph*{Dirichlet boundary condition.}
For systems whose edge properties correspond to Dirichlet boundary conditions,
the number of available edge channels fixed by the boundary coincides with the
number of occupied Landau-level families,
\[
\nu_c=\nu_n \equiv n_{\rm max}.
\]
The Hall resistivity therefore reduces to
\begin{equation}\label{IQHE_exp}
\rho_{xy}=\frac{h}{e^2}\frac{1}{\nu_p},
\qquad
\nu_p=1,2,\ldots,n_{\rm max},
\end{equation}
which reproduces the standard integer quantum Hall effect (IQHE) sequence.
This case therefore serves as the natural reference limit in which boundary
quantization reproduces the conventional integer quantum Hall response.

We now examine how Neumann and Robin boundary conditions modify the edge
spectrum. In contrast to the Dirichlet case, the boundary-induced change in
edge-channel multiplicity generates additional admissible filling factors
and leads naturally to the fractional plateaus observed experimentally.

\paragraph*{Neumann boundary condition.}
For systems described by Neumann boundary conditions, the number of available
edge channels is increased by one,
\[
\nu_c=\nu_n+1.
\]
The effective filling factor then becomes
\[
\nu_{\rm eff}=\frac{\nu_p\,n_{\rm max}}{n_{\rm max}+1},
\qquad
\nu_p=1,2,\ldots,n_{\rm max}+1,
\]
reflecting the presence of an additional boundary-allowed edge channel that
can remain partially populated near the Fermi level.
This construction generates a hierarchy of effective filling-factor
sequences for all values of $n_{\rm max}$. For clarity, we list here the
lowest-order cases explicitly:
\begin{equation}\label{expliciteven}
\begin{array}{rcll}
n_{\rm max}=1 &\Rightarrow&
\nu_{\rm eff}=\dfrac{\nu_p}{2}
&=\left\{\dfrac{1}{2},\,1\right\},\\[8pt]
n_{\rm max}=2 &\Rightarrow&
\nu_{\rm eff}=\dfrac{2\nu_p}{3}
&=\left\{\dfrac{2}{3},\,\dfrac{4}{3},\,2\right\},\\[8pt]
n_{\rm max}=3 &\Rightarrow&
\nu_{\rm eff}=\dfrac{3\nu_p}{4}
&=\left\{\dfrac{3}{4},\,\dfrac{3}{2},\,\dfrac{9}{4},\,3\right\},\\[8pt]
n_{\rm max}=4 &\Rightarrow&
\nu_{\rm eff}=\dfrac{4\nu_p}{5}
&=\left\{\dfrac{4}{5},\,\dfrac{8}{5},\,\dots,\,4\right\}.
\end{array}
\end{equation}
Higher values of $n_{\rm max}$ extend this sequence systematically,
generating additional admissible fractions according to the same
boundary-quantization rule.

\paragraph*{Robin boundary condition.}
For systems whose edges are described by Robin boundary conditions, the number
of available edge channels is increased by two,
\[
\nu_c=\nu_n+2.
\]
The effective filling factor is now
\[
\nu_{\rm eff}=\frac{\nu_p\,n_{\rm max}}{n_{\rm max}+2},
\qquad
\nu_p=1,2,\ldots,n_{\rm max}+2,
\]
corresponding to the presence of two additional boundary-allowed edge channels,
one at each edge, whose occupation is controlled by the same energetic pinning
mechanism.
This construction generates a hierarchy of effective filling-factor
sequences for all values of $n_{\rm max}$ under Robin boundary conditions.
For clarity, we list here several representative low-order cases:
\begin{equation}\label{explicitodd}
\begin{array}{rcll}
n_{\rm max}=1 &\Rightarrow&
\nu_{\rm eff}=\dfrac{\nu_p}{3}
&=\left\{\dfrac{1}{3},\,\dfrac{2}{3},\,1\right\},\\[8pt]
n_{\rm max}=2 &\Rightarrow&
\nu_{\rm eff}=\dfrac{2\nu_p}{4}
&=\left\{\dfrac{1}{2},\,1,\,\dfrac{3}{2},\,2\right\},\\[8pt]
n_{\rm max}=3 &\Rightarrow&
\nu_{\rm eff}=\dfrac{3\nu_p}{5}
&=\left\{\dfrac{3}{5},\,\dfrac{6}{5},\,\dots,\,3\right\},\\[8pt]
n_{\rm max}=4 &\Rightarrow&
\nu_{\rm eff}=\dfrac{4\nu_p}{6}
&=\left\{\dfrac{2}{3},\,\dfrac{4}{3},\,2,\,\dots,\,4\right\},\\[8pt]
n_{\rm max}=5 &\Rightarrow&
\nu_{\rm eff}=\dfrac{5\nu_p}{7}
&=\left\{\dfrac{5}{7},\,\dfrac{10}{7},\,\dots,\,5\right\}.
\end{array}
\end{equation}
Higher values of $n_{\rm max}$ extend this hierarchy systematically,
producing additional admissible fractions according to the same
boundary-quantization rule.
In particular, the characteristic fractional values
\[
\nu_{\rm eff}=\frac{1}{3},\,\frac{2}{3},\,\frac{3}{5},\,\frac{4}{5},\,\frac{5}{7},\ldots
\]
emerge naturally from the same finite-system quantization rules, without
requiring electron-electron correlations, composite fermions, or
disorder-induced localization as primary ingredients of the quantization
mechanism.

This does not exclude the presence or importance of interactions and disorder
in real samples; rather, it shows that the observed filling-factor hierarchy
can already be generated at the level of boundary-controlled edge spectra,
with additional many-body and scattering effects acting on a structure that is
intrinsically established by the boundary quantization.

Within this unified framework, the integer and fractional quantum Hall effects
emerge as complementary manifestations of a common microscopic organizing
mechanism: the discrete, boundary-dependent spectrum of edge states imposed by
lateral confinement in a finite two-dimensional electron system.
\begin{figure*}[t]
\centering
\includegraphics[width=16.5cm]{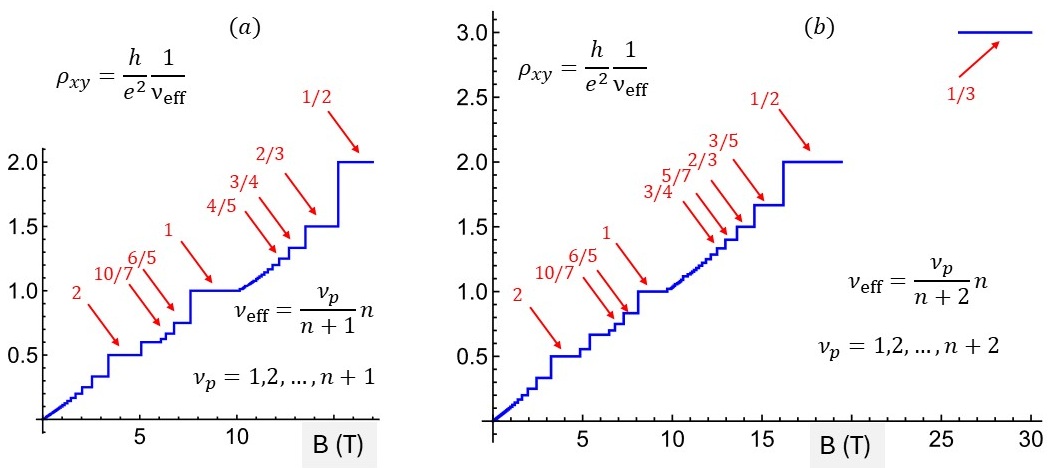}
\caption{
Predicted Hall resistance $\rho_{xy}=(h/e^2)/\nu_{\rm eff}$ obtained from
boundary-quantized edge transport for (a) Neumann and (b) Robin boundary
conditions as a function of magnetic field.
The effective filling factors follow the sequences
$\nu_{\rm eff}=\nu_p/(n+1)$ for Neumann and
$\nu_{\rm eff}=\nu_p/(n+2)$ for Robin boundaries, with
$\nu_p=1,2,\ldots,n+1$ and $\nu_p=1,2,\ldots,n+2$, respectively.
The principal fractional plateaus are indicated in red.
Plateau widths arise directly from the density of allowed filling
factors generated by the boundary quantization.
For the same magnetic-field interval, the Neumann hierarchy produces
slightly broader plateaus, while the Robin case yields a denser but more
closely spaced sequence extending toward smaller fractional values.
}
\label{fig:NeumannRobinHall}
\end{figure*}
\paragraph*{Integer and fractional plateaus from boundary quantization}

To connect the boundary-controlled hierarchy with experimental observations,
we now relate the effective filling-factor sequences to the magnetic-field
dependence of the Hall resistance.
Dirichlet boundary conditions reproduce the conventional integer sequence,
while Neumann and Robin boundary conditions generate fractional structures
through their modified edge-channel multiplicities.

A key ingredient in this comparison is the relation between the highest
occupied Landau-level index and the total energy $E_T$,
\begin{equation}\label{lfloor}
\nu_n = n_{\rm max} =
\left\lfloor \frac{E_T - \hbar\omega_c/2}{\hbar\omega_c} \right\rfloor ,
\end{equation}
where $\lfloor x \rfloor$ denotes the integer part of $x$.
In contrast to conventional Landau-level counting, this expression
explicitly incorporates the zero-point contribution.

Microscopically, $E_T$ corresponds to the discrete set of edge-state energies
$E_{nj}=K_{nj}+U_{nj}$ introduced above.
It includes the Fermi energy $E_F$ of the transmitted electrons, fixed by the
reservoir chemical potentials, together with the Landau zero-point energy
$\hbar\omega_c/2$.
The inclusion of this contribution is essential for reproducing the correct
low-field behavior of the Hall resistance. Neglecting this zero-point contribution leads to an incorrect low-field curvature of $\rho_{xy}(B)$, replacing the experimentally observed quasi-linear onset by a parabolic dependence.
Accordingly, one may write $E_T \simeq E_F + \hbar\omega_c/2$, so that $\nu_n$
counts the number of fully occupied Landau levels lying below $E_T$, changing
stepwise as the magnetic field varies.

Substituting Eq.~(\ref{lfloor}) into the general expression for the Hall
resistivity gives
\begin{equation}\label{Hallreslfloor}
\rho_{xy} = \frac{h}{e^2}
\frac{1}{\displaystyle \frac{\nu_p}{\nu_c}\left\lfloor\frac{2\pi m^* E_F}{h e B} \right\rfloor},
\end{equation}
where \(m^*\) is the effective electron mass.
This relation does not introduce a new quantization rule; it connects the
boundary-determined hierarchy of edge states with the magnetic-field–dependent
indexing that governs transport measurements.
In the Dirichlet limit, \(\nu_c=n_{\rm max}\) and therefore
\(\nu_{\rm eff}=\nu_p\). In this case Eq.~(\ref{Hallreslfloor}) should not be
interpreted as a direct translation into bulk Landau degeneracy. Instead, it
encodes the discrete set of transmitting edge states available up to the cutoff
\(n_{\rm max}\), i.e., all branches with \(n \leq n_{\rm max}\). The
experimentally observed plateau sequence then emerges as the magnetic field
sequentially reveals this hierarchy when individual edge branches intersect the
Fermi level and depopulate.
Quantization therefore originates at the boundary level, while the Landau-level
description emerges as an effective macroscopic representation.
The resulting magnetic-field dependence reproduces the quasi-linear sequence
of Hall plateaus observed experimentally (Fig.~\ref{graphsIQHE}) and clarifies
the distinction between the discrete edge-state energies $E_{nj}$ and the
macroscopic Fermi energy $E_F$ set by the reservoirs.

In Fig.~\ref{graphsIQHE}(a,b) we compare the predicted Hall resistance with
experimental results for integer plateaus in GaInAs/InP~\cite{Koch1993} and
GaAs/(AlGa)As~\cite{WeisVonKlitzing}, using the corresponding effective masses
and Fermi energies indicated in the figure.
The theoretical curves, obtained assuming Dirichlet boundary conditions, are
overlaid on the experimental data and show excellent agreement across the full
magnetic-field range.

It is well established that, as the magnetic field increases and the cyclotron
energy $\hbar\omega_c$ grows, the bulk Landau levels successively cross the
Fermi energy.
Within the present finite-system framework, however, a subset of low-energy
edge branches can remain below the Fermi level even after the corresponding
bulk Landau level has depopulated, as illustrated in
Fig.~\ref{kineticenergies}.
Fractional quantization therefore arises from the systematic survival of edge
states that maintain current continuity across Landau-level transitions,
without requiring electron--electron correlations as a primary quantization
mechanism.

Figure~\ref{fig:NeumannRobinHall} shows the Hall-resistance staircase
predicted when the edge spectrum is determined solely by boundary
conditions. In both panels the plateau structure is not imposed
phenomenologically but emerges from the discrete hierarchy of effective
filling factors allowed by boundary-quantized edge channels.

In the Neumann case [Fig.~\ref{fig:NeumannRobinHall}(a)], the additional
boundary-allowed channel increases the multiplicity of edge states,
producing fractional plateaus such as $2/3$, $3/4$, $4/5$, and $6/5$.
Because the admissible fractions are more sparsely distributed, the
magnetic-field intervals over which $\rho_{xy}$ remains constant are
comparatively broad.

For Robin boundary conditions [Fig.~\ref{fig:NeumannRobinHall}(b)], two
additional boundary-controlled channels generate a richer hierarchy that
extends to smaller fractions, including $1/3$, $3/5$, and $5/7$.
The higher density of admissible filling factors leads to a finer
staircase and correspondingly narrower plateaus within the same field
range.

Plateau widths therefore follow directly from the arithmetic structure of
the boundary-generated filling-factor sequences.
Above $\nu=1$, the admissible fractions accumulate toward unity as
$(n+1)/n$ and $(n+2)/n$ for Neumann and Robin boundaries, respectively,
producing closely spaced steps.
Below $\nu=1$, the density of allowed fractions is comparatively lower and
the plateaus become more widely separated.
This intrinsic asymmetry in the distribution of admissible filling factors
determines the characteristic widths of the Hall plateaus within the
boundary-quantized spectrum.

Disorder and inelastic processes do not create these plateaus; rather,
they broaden and round the staircase already established by the boundary
quantization and by the arithmetic organization of the effective
filling-factor hierarchy.

Boundary conditions discretize both the guiding-center positions $y_0$
and the longitudinal momenta of edge states.
Within this framework, the Neumann condition may be viewed as a limiting
case of the more general Robin boundary condition.
More broadly, boundary confinement introduces a fine structure in the
longitudinal edge momentum, analogous to a hyperfine splitting, whose
magnitude sets the characteristic energy scale of fractional
quantization.

\begin{figure*}[hbt]
\begin{center}
\includegraphics[width=16.5cm]{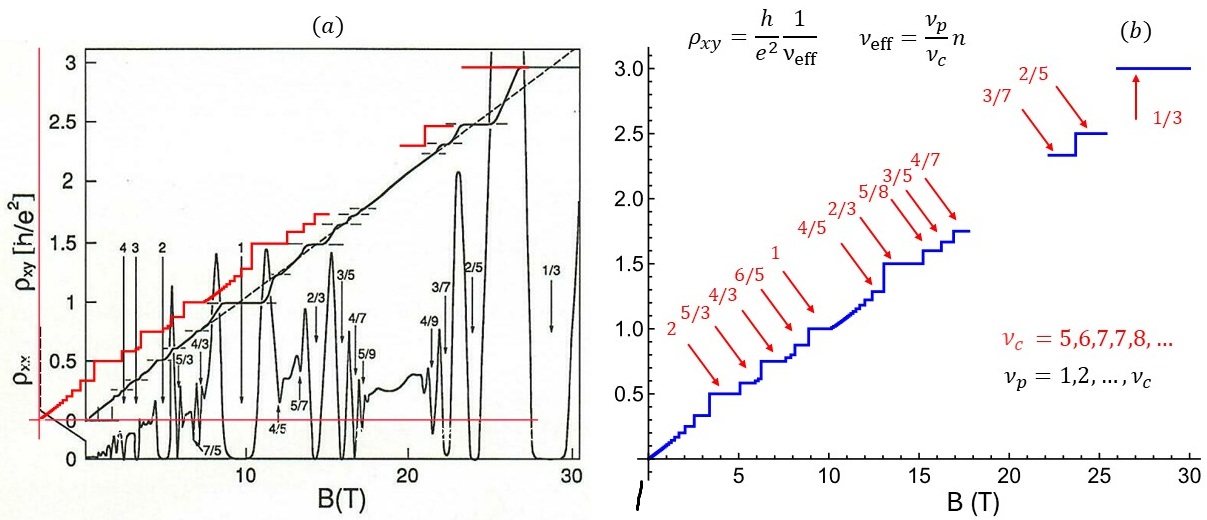}
\caption{
Experimental and theoretical fractional Hall resistance based on boundary-quantized edge states with parity breaking.
(a) Experimental Hall resistance $\rho_{xy}/(h/e^2)$ measured by St\"ormer \textit{et al.}~\cite{Stormer1992} (reproduced with permission from Elsevier).
(b) Theoretical Hall resistance calculated within the present framework using Neumann boundary conditions and a weak parity-breaking parameter $b=0.004$, for the Fermi energy $E_F=0.001125$eV.
The theoretical prediction (red curve) is overlaid on the experimental data to facilitate direct comparison.
The prominent high-field fractional plateaus at $2/5$ and $3/7$ observed experimentally are reproduced by the theory.
}\label{graphsFQHE2}
\end{center}
\end{figure*}

We have shown that boundary conditions discretize both the guiding-center
positions $y_0$ and the longitudinal momenta of edge states.
More generally, this framework demonstrates that lateral confinement
introduces a fine structure in the longitudinal edge momentum, analogous
to a hyperfine splitting, whose magnitude sets the characteristic energy
scale of fractional quantization.

Within this boundary-controlled picture, the integer hierarchy and a first
set of fractional plateaus arise directly from the discrete spectrum of
edge states.
However, experiments at higher magnetic fields reveal additional fractional
structures that cannot be accounted for by boundary quantization alone.
To describe these regimes, we now include the parity-breaking mechanism
introduced at the end of the previous section.

\paragraph*{Fractional plateaus from parity-enhanced multiplicity}

Retaining a small fraction of the Hall voltage as a parity-breaking
perturbation increases the number of low-energy edge states, particularly
at small Landau indices.
When combined with a Neumann and Robin boundary condition, this perturbation further
enhances the multiplicity of boundary-supported channels, producing extended
families of effective filling factors and enabling the emergence of
additional high-field fractional plateaus.

To illustrate the robustness of this mechanism, we list two representative
multiplicity patterns corresponding to nearby regions of the $(b,L_y)$
parameter space. Both generate closely related fractional hierarchies and
reproduce the principal experimentally observed plateaus.

\begin{table}[hbt]
\centering
\caption{Effective filling-factor sequences generated by parity-enhanced edge
multiplicities for the representative sequence $(7,5,6,7,8,\ldots)$.}
\label{tab:ParitySequencesA}
\begin{tabular}{l c c c}
\hline\hline
 & \; \, & \;  & \,
Effective filling factors \\  $n$ \,&$\,\nu_c$ &$\nu_n$ & $\nu_{\rm eff}=(\nu_p/\nu_c)\nu_n$ \\[6pt]
\hline
$1$ & $7$ & $1$ &
$\left\{1/7, 2/7, 3/7, 4/7, 5/7, 6/7, 1\right\}$ \\[10pt]

$2$ & $5$ & $2$ &
$\left\{2/5, 4/5, 6/5, 8/5, 2\right\}$ \\[10pt]

$3$ & $6$ & $3$ &
$\left\{1/2, 1, 3/2, 2, 5/2, 3\right\}$ \\[10pt]

$4$ & $7$ & $4$ &
$\left\{4/7, 8/7, 12/7, 16/7, 20/7, 24/7, 4\right\}$ \\[10pt]

$5$ & $8$ & $5$ &
$\left\{5/8, 10/8, 15/8, 20/8, 25/8, 30/8, 35/8, 5\right\}$ \\[6pt]
\hline\hline
\end{tabular}
\end{table}

\begin{table}[hbt]
\centering
\caption{Effective filling-factor sequences generated by parity-enhanced edge
multiplicities for the representative sequence $(5,6,7,7,8,\ldots)$,
used for comparison with experimental fractional hierarchies.}
\label{tab:ParitySequencesB}
\begin{tabular}{l c c c}
\hline\hline
 \;& \; \, & \;  & \,
Effective filling factors \\  $n$ \;\, & \,$ \nu_c$& \,$\nu_n$& $\nu_{\rm eff}=(\nu_p/\nu_c)\nu_n$ \\[6pt]
\hline
$1$ & $5$ & $1$ &
$\left\{1/5, 2/5, 3/5, 4/5, 1\right\}$ \\[10pt]

$2$ & $6$ & $2$ &
$\left\{1/3, 2/3, 1, 4/3, 5/3, 2\right\}$ \\[10pt]

$3$ & $7$ & $3$ &
$\left\{3/7, 6/7, 9/7, 12/7, 15/7, 18/7, 3\right\}$ \\[10pt]

$4$ & $7$ & $4$ &
$\left\{4/7, 8/7, 12/7, 16/7, 20/7, 24/7, 4\right\}$ \\[10pt]

$5$ & $8$ & $5$ &
$\left\{5/8, 10/8, 15/8, 20/8, 25/8, 30/8, 35/8, 5\right\}$ \\[6pt]
\hline\hline
\end{tabular}
\end{table}

The second sequence listed above is used in the transport calculation shown
in Fig.~\ref{graphsFQHE2}, where the resulting Hall resistance is compared
directly with experimental data. Panel~(a) displays the measured Hall
resistance, while panel~(b) shows the theoretical prediction obtained from
boundary quantization with a weak parity-breaking interaction.
In addition to the low-field plateaus, the high-field plateaus corresponding
to fractions $3/7$ and $2/5$ are clearly reproduced.
A perturbative analysis of zero counting in the presence of parity breaking
is presented in Appendix~E.

The theoretical predictions reproduce the experimentally observed integer
and fractional plateaus at low, intermediate, and high magnetic fields with
remarkable accuracy, without invoking disorder, spin splitting, or
phenomenological localization mechanisms, and within a single, unified
boundary-based framework.

\begin{figure*}[t]
\centering
\includegraphics[width=10.5cm]{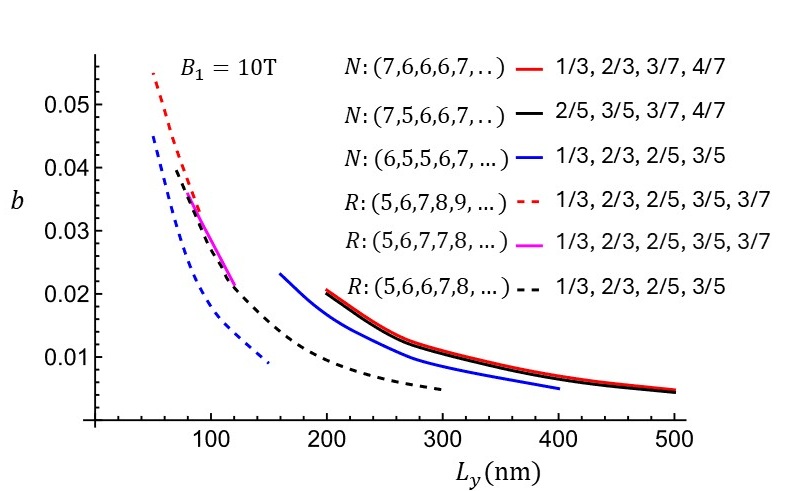}
\caption{
Neumann and Robin multiplicity sequences and their geometric loci as
functions of $b$ and $L_y$.
Given the magnetic field $B_1$ at which the plateau for $n=1$ occurs, and
choosing $b$ and $L_y$, imposing the boundary conditions on the Landau
wave packets $\eta_y$, and counting the number of zeros as described in
the text and Appendix~D, one obtains a sequence of effective filling
factors.
Searching for the same sequence in neighboring points of the parameter
space $(L_y,b)$ determines the geometric locus associated with each
sequence of filling factors.
The figure shows the loci corresponding to sequences responsible for
fractional plateaus such as $1/3$, $2/3$, $2/5$, $3/5$, $3/7$, etc., at
high magnetic fields.
}
\label{GeometricLoci}
\end{figure*}

\paragraph*{Geometric loci of boundary- and parity-generated sequences}

The multiplicity sequences produced by boundary quantization and their
modification under parity breaking can be represented geometrically in the
parameter space defined by the parity-breaking strength $b$ and the lateral
confinement scale $L_y$.
Figure~\ref{GeometricLoci} shows the loci associated with representative
fractional sequences obtained by imposing the boundary conditions on the
Landau wave packets and counting the corresponding zeros, as described in
Appendix~D.

For a fixed magnetic field, the allowed edge spectra are determined jointly
by the confinement length $L_y$ and the parity-breaking parameter $b$.
Each point in the $(b,L_y)$ plane corresponds to a specific multiplicity of
low-energy edge channels and therefore to a well-defined sequence of effective
filling factors.
The geometric curves shown in Fig.~\ref{GeometricLoci} identify the regions
of parameter space where a given sequence is realized and remains stable
under small variations of $b$ and $L_y$.

In this representation, fractional plateaus do not arise as isolated numerical
coincidences but as organized families of edge-state spectra.
Sequences responsible for plateaus such as $1/3$, $2/3$, $2/5$, $3/5$, and
$3/7$ occupy distinct but partially overlapping domains of the $(b,L_y)$
space.
Transitions between these domains correspond to changes in the number of
boundary-supported edge branches and therefore to reorganizations of the
effective filling-factor hierarchy.

The geometric picture also clarifies the complementary roles of boundary
quantization and parity breaking.
In the limit $b \rightarrow 0$, the loci reduce to those generated purely by
Neumann and Robin boundary conditions, and the corresponding sequences
coincide with the boundary-only hierarchy discussed above.
As $b$ increases, the loci deform and new families of fractional sequences
appear, reflecting the enhanced multiplicity of low-energy edge states.
Parity breaking therefore acts as a continuous control parameter that shifts
the system between boundary-dominated and parity-enhanced regimes without
altering the underlying Landau quantization.

This geometric organization explains the coexistence of several fractional
plateaus within the same magnetic-field interval.
For example, the $1/3$ plateau can occur both in wide samples with weak
parity breaking and in narrower samples where parity-enhanced edge
multiplicities are present, while fractions such as $2/5$ and $3/7$ emerge
predominantly in regions where parity breaking increases the density of
admissible edge states.
The observed hierarchy of fractional plateaus is thus mapped onto a
structured landscape in the $(b,L_y)$ plane.

When the sequences identified in Fig.~\ref{GeometricLoci} are incorporated
into the transport calculation, the resulting Hall resistance reproduces
the experimental high-field fractional plateaus shown in
Fig.~\ref{graphsFQHE2}.
The geometric loci therefore provide a unifying interpretation: they link
the microscopic zero-counting procedure, the boundary-controlled edge
multiplicity, and the macroscopic transport signatures observed in the
fractional quantum Hall regime.

\section{Discussion and Conclusions}

The analysis developed in this work establishes a unified microscopic picture of
quantized Hall transport in laterally confined two-dimensional electron systems
by making explicit the role of boundaries in shaping the edge spectrum.
Rather than treating edge states as secondary to bulk topology or as
phenomenological transport channels, the present analysis shows that
physically imposed boundary conditions directly determine the discrete
structure, multiplicity, and occupancy of chiral edge modes.
Transport quantization then follows from the propagation of these
boundary-quantized states.

Within this picture, the quantum Hall effect emerges from the interplay of three
interconnected elements. Bulk Landau quantization defines the existence and
chirality of edge modes. Boundary conditions discretize guiding-center positions
and longitudinal momenta, generating a structured hierarchy of edge channels.
Phase-coherent transport through these channels produces the macroscopic Hall
plateaus observed experimentally. This hierarchy provides a direct microscopic
bridge between Landau quantization and the transport spectrum in real devices.

A central outcome of the theory is that physically consistent boundary
conditions -- Dirichlet, Neumann, and mixed (Robin-type) -- lead to closely related
edge localization patterns while differing in their allowed degeneracies and
channel multiplicities. Boundary quantization alone accounts for the dominant
integer Hall sequence and for several structured fractional features at low and
intermediate magnetic fields. In this regime, quantization reflects the discrete
spectrum of boundary-supported chiral modes rather than the global filling of
bulk Landau levels alone.

At higher magnetic fields, a complete description requires incorporating a weak
parity-breaking contribution associated with Hall-induced asymmetry,
structural imbalance, tunneling, and unequal edge transmission. This controlled
symmetry breaking reorganizes the low-energy edge spectrum without altering the
underlying Landau-level ordering. When combined with Neumann or Robin boundary
conditions, it enhances the multiplicity of low-index edge states and generates
extended families of fractional filling factors. The fractional hierarchy thus
arises from a boundary-driven reorganization of transmitting channels rather
than from a modification of the bulk quantization itself.

From this perspective, bulk topology guarantees the existence and robustness of
chiral edge modes, but the quantitative spectrum and multiplicity of those modes
are determined by confinement and boundary physics. The effective filling factors
measured experimentally correspond to the number of boundary-allowed and
energetically populated chiral channels. Disorder and interactions, while
important in real materials, act primarily as modifiers of an already quantized
boundary spectrum rather than as the primary origin of quantization.

A key microscopic feature revealed by this analysis is the persistence of
low-energy edge branches below the Fermi level even after the associated bulk
Landau levels have crossed it. These long-lived boundary states maintain the
continuity of the chiral current as the magnetic field varies and provide a
geometric mechanism for the formation and evolution of Hall plateaus.
Broad plateaus correspond to dense families of persistent edge modes, whereas
narrower structures arise from higher-order branches with larger energy
separations. The cascade of fractional plateaus beyond the first integer plateau
therefore reflects the sequential survival and depopulation of
boundary-quantized channels.

Within this unified microscopic description, the integer and fractional quantum
Hall effects appear as complementary manifestations of the same underlying
mechanism: confinement-induced quantization of edge spectra refined by weak
parity breaking. The quantized Hall resistance,
\begin{equation}
  \rho_{xy}=\frac{h}{e^2\nu_{\rm eff}}, \qquad
  \nu_{\rm eff}=\frac{\nu_p}{\nu_c}\,\nu_n,
\end{equation}
directly encodes the boundary-induced discretization of the transmitting edge
structure and provides a compact analytical connection between microscopic
spectra and macroscopic transport.

Beyond unifying the integer and fractional regimes, this approach establishes a
general microscopic foundation for quantized transport in confined electronic
systems. Because it relies only on geometric confinement, boundary-induced spectral
discretization, and coherent edge propagation, its principles extend naturally
to a broad class of materials in which boundary symmetries and confinement play
a decisive role, including graphene, semiconductor heterostructures,
topological insulators, and moir\'e systems.

More broadly, the present results clarify the relationship between topology and
microscopic confinement in quantum Hall physics. Topological arguments ensure
the robustness and chirality of edge transport, whereas boundary conditions
determine the spectral structure through which quantization is realized.
By explicitly connecting boundary physics, edge spectra, and transport,
the theory bridges the gap between bulk invariants and the experimentally
observed hierarchy of Hall plateaus.

This boundary-centered perspective opens several directions for future work.
Extensions to interacting systems, nonuniform confinement, and time-dependent
driving may reveal additional spectral reorganizations of edge channels.
Likewise, engineering boundary conditions through electrostatic gating or
nanostructuring offers a route to tailoring edge multiplicities and transport
quantization in designed quantum devices. The analysis presented here therefore
provides not only a unified interpretation of quantum Hall phenomena but also a
basis for exploring boundary-controlled quantization in a wide range of quantum
materials and mesoscopic systems.

\appendix
\section{Robin boundary condition and curvature constraint for Landau wave functions}

We consider the general Robin boundary condition imposed at the sample edges,
\begin{equation}\label{RobinDef}
\left.
\left(
a\,\eta(y)+\frac{\partial \eta(y)}{\partial y}
\right)
\right|_{y=\pm L_y/2}
=0,
\end{equation}
where $a$ is a real parameter specifying the effective boundary constraint.

A basic physical requirement for a laterally confined system is the absence of
probability flow normal to the boundary. The normal component of the quantum
probability current density is
\begin{equation}\label{NormalCurr}
j_n \propto
\Im\!\left[
\eta^*(y)\,\frac{\partial \eta(y)}{\partial y}
\right].
\end{equation}
Since the transverse Landau functions $\eta(y)$ can be chosen real in the
present gauge, the Robin condition (\ref{RobinDef}) guarantees that $j_n=0$ at
$y=\pm L_y/2$. Robin boundary conditions are therefore fully consistent with
confinement at the sample edges.
\begin{figure*}[hbt]
\begin{center}
\includegraphics[width=17cm]{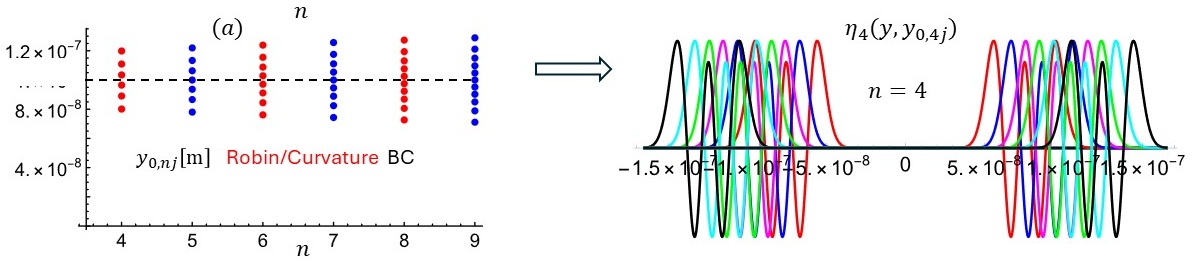}
\caption{
Guiding-center quantization under Robin boundary conditions.
(a) Discrete guiding-center positions $y_{0,nj}$ selected by the Robin
constraint, Eq.~(\ref{BC_Robin}).
Right column: corresponding Landau wave packets evaluated at these quantized
guiding centers for $n=4$.
Robin boundary conditions generate edge-localized branches closely analogous to the Dirichlet and Neumann cases (Fig.~\ref{yonQuant}), but with two additional boundary-admissible solutions appearing near the inflection-point region of the Gaussian envelope of the Landau wave packet.
Parameters: $B=15$~T and $L_y=0.2~\mu$m.
}
\label{Robinyon}
\end{center}
\end{figure*}

To express Eq.~(\ref{RobinDef}) in the Landau basis, we introduce the
dimensionless coordinate $\xi$ defined in Eq.~(\ref{defxi}), with boundary
values $\xi_\pm=\xi|_{y=\pm L_y/2}$. Using the Gauss--Hermite form of Landau
wave functions, the Robin constraint becomes
\begin{equation}\label{RobinHermite}
2n\,H_{n-1}(\xi_\pm)
=
(\xi_\pm-a)\,H_n(\xi_\pm),
\end{equation}
and, by iteration,
\begin{equation}\label{RobinHermite2}
2n\,2(n-1)\,H_{n-2}(\xi_\pm)
=
(\xi_\pm-a)^2\,H_n(\xi_\pm).
\end{equation}

The second derivative of the transverse wave function can be written as
\begin{eqnarray}\label{SecondDerivGeneral}
\frac{\partial^2 \eta(y)}{\partial y^2}
=
\frac{e^{-\xi^2/2}}{\ell_B^2}
\Bigl[
(\xi^2-1)\,H_n(\xi)
-2\xi\,2n\,H_{n-1}(\xi)\crcr
+2n\,2(n-1)\,H_{n-2}(\xi)
\Bigr].
\end{eqnarray}

Evaluating Eq.~(\ref{SecondDerivGeneral}) at the boundaries and substituting
Eqs.~(\ref{RobinHermite})--(\ref{RobinHermite2}) yields
\begin{equation}\label{SecondDerivEdge}
\left.
\frac{\partial^2 \eta(y)}{\partial y^2}
\right|_{y=\pm L_y/2}
=
\frac{e^{-\xi_\pm^2/2}}{\ell_B^2}\,(a^2-1)\,H_n(\xi_\pm).
\end{equation}

Equation (\ref{SecondDerivEdge}) shows that Robin boundary conditions impose a
well-defined curvature of the transverse wave function at the sample edges.
In particular, for the special values $a=\pm1$ one obtains
\[
\eta''(\pm L_y/2)=0 ,
\]
independently of the Landau index $n$ (except at accidental nodes
$H_n(\xi_\pm)=0$).

In this work we focus on the special Robin boundary condition with $a=1$.
This ``curvature-matching'' constraint  provides a physically
transparent interpolation between Dirichlet and Neumann conditions while
preserving zero normal current at the sample edges.

\begin{figure*}[hbt]
\begin{center}
\includegraphics[width=15cm]{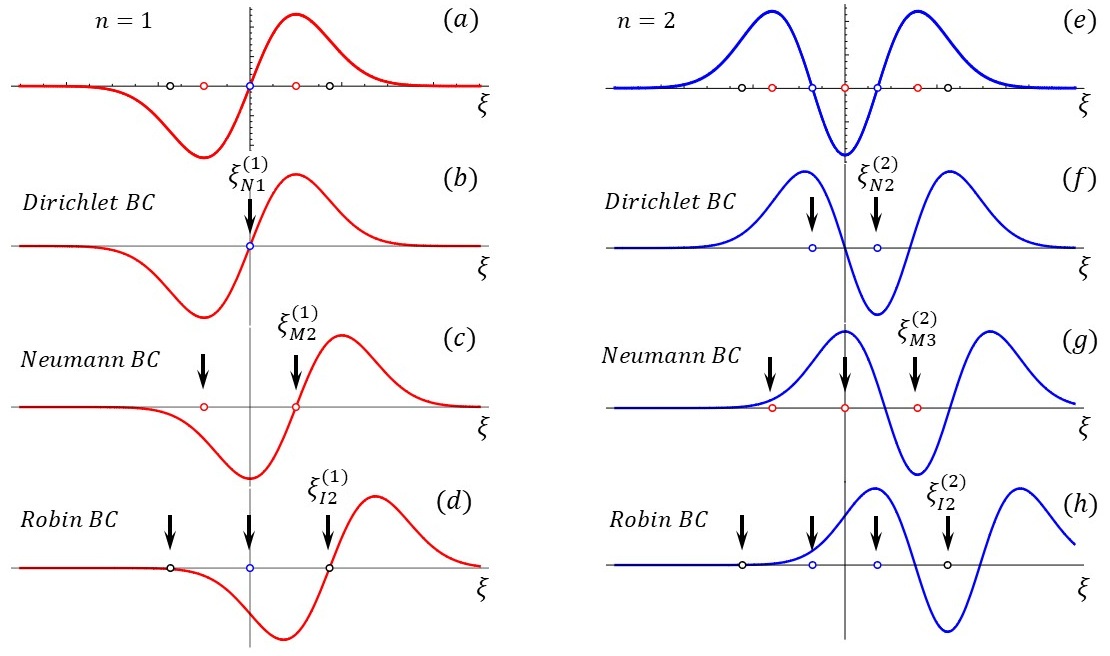}
\caption{Landau wave packets and the geometric meaning of Dirichlet, Neumann,
and Robin boundary conditions.
Left column: $n=1$; right column: $n=2$.
Top panels show the Gauss--Hermite Landau wave packet
$\eta_n(\xi)\propto e^{-\xi^2/2}H_n(\xi)$ together with its characteristic
critical points: nodes (Dirichlet zeros, blue), extrema (Neumann zeros,  red),
and outer inflection points that generate additional Robin-admissible solutions
(black).
Second row: wave packets evaluated at the guiding-center positions ($\xi_{N1}$, $\xi_{N2}$), 
admitted by Dirichlet boundary conditions at the sample edge, illustrating that
Dirichlet constraints pin guiding centers to the nodes of $\eta_n$.
Third row: corresponding Neumann-admissible guiding centers ($\xi_{M1}$, $\xi_{M2}$, $\xi_{M3}$), which coincide with
the extrema of $\eta_n$.
Bottom row: Robin-admissible guiding centers for the special condition
$\pm\eta_n+\eta_n'=0$, which includes both node-like solutions and two additional
solutions ($\xi_{I1}$ and $\xi_{I2}$), near the outer inflection (turning-point) region of the wave packet.}
\label{LWPatDNRZ}
\end{center}
\end{figure*}
\section{Physical meaning of Dirichlet, Neumann, and Robin boundary conditions}

In conventional one-dimensional quantum wells, Dirichlet, Neumann, and Robin
boundary conditions are often interpreted as describing ``hard'' or ``soft''
confining walls that suppress the wavefunction at the boundaries and produce
strong level repulsion.
For Landau quantization in a magnetic field, this interpretation is
fundamentally misleading.
The boundary conditions do not confine the wavefunction in the usual sense,
do not truncate its Gaussian tails outside the nominal sample region, and do
not generate the infinite-well type of level spacing.
Instead, their essential role is to quantize the admissible guiding-center
positions of displaced Landau wave packets.

\vspace{0.15cm}
\noindent\textbf{1.\ Landau wave packets and boundary zeros.}
To clarify the microscopic action of boundary conditions, it is useful to
express the transverse Landau eigenfunctions in terms of the dimensionless
variable $\xi$ defined in Eq.~(\ref{defxi}),
\begin{equation}
\eta_n(\xi)\propto e^{-\xi^2/2}\,H_n(\xi),
\end{equation}
which has the form of a Gauss--Hermite wave packet.
For each Landau index $n$, this structure possesses $n$ nodes, $n+1$ local
extrema, and two outer inflection points associated with the Gaussian envelope.

These characteristic points acquire a direct boundary interpretation.
Dirichlet boundary conditions,
\begin{equation}
\eta_n(\pm L_y/2)=0,
\end{equation}
select guiding centers located at the nodes of $\eta_n$, i.e., at the $n$ real
zeros of the Hermite polynomial $H_n$.
Neumann boundary conditions,
\begin{equation}
\eta_n'(\pm L_y/2)=0,
\end{equation}
instead select guiding centers at the extrema of the Gauss--Hermite packet,
corresponding to the $n+1$ real zeros of $\eta_n'$.

More generally, Robin boundary conditions take the form
\begin{equation}
\left.(a\,\eta_n+\eta_n')\right|_{y=\pm L_y/2}=0,
\end{equation}
where $a$ is a real boundary parameter.
In this work we focus on the curvature-matching choice $a=\pm 1$, which is
equivalent to the second-derivative constraint discussed in the preceding
subsection.
For this special case, the Robin condition yields the Dirichlet node spectrum
together with two additional admissible solutions located near the outer
inflection region of the Landau wave packet, producing $n+2$ guiding-center
roots.

\begin{figure*}[hbt]
\begin{center}
\includegraphics[width=16.5cm]{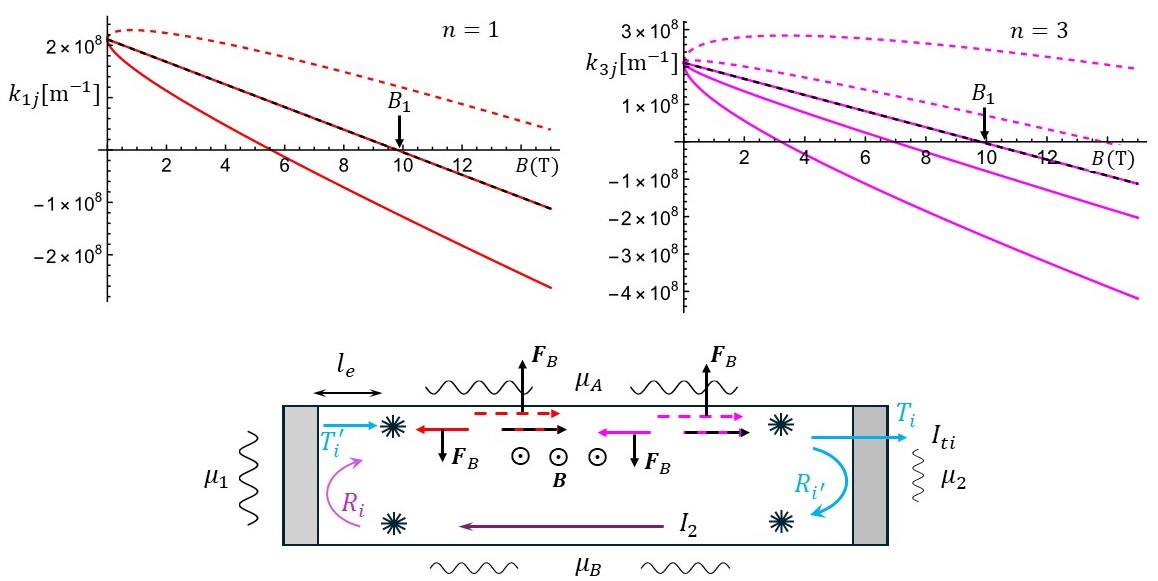}
\caption{
Evolution of the longitudinal momenta $k_{x,nj}$ of boundary-quantized edge
channels with magnetic field $B$.
Solid colored curves show representative branches for different Landau indices
$n$.
Dotted curves denote the \emph{pinned} edge states that remain localized near
the upper boundary and continue to carry the chiral current as $B$ increases.
Dashed black curves (overlaid in color for each $n$) correspond to the
$\xi_{nj}=0$ solutions, which are purely kinetic and provide the reference
momentum scale for the localized channels.
A change in the sign of $k_{x,nj}$ indicates that the corresponding state
exchanges from one edge to the opposite one, reversing its propagation direction
on a given boundary while preserving the overall chirality of the system.
The field $B_1=9.8$\,T marks the onset of the $n=1$ plateau.
Parameters: $m^*=0.067\,m_e$, $E_F=(3/2)\hbar\omega_{c1}$ with
$\omega_{c1}=eB_1/m^*$, and effective stripe width $L_y=28.3896$\,nm at $B_1$.
}
\label{EdgeStatesMomenta}
\end{center}
\end{figure*}
These correspondences are illustrated in Fig.~\ref{LWPatDNRZ}, where the nodes
(Dirichlet), extrema (Neumann), and inflection-related roots (Robin) are marked
explicitly for the lowest Landau levels.
Evaluating $\eta_n$ at these boundary-selected points shows that the guiding
center is discretized precisely by the zeros of the imposed boundary
constraint.

\vspace{0.15cm}
\noindent\textbf{2.\ Guiding-center quantization without confinement.}
The Landau eigenstates in a finite Hall bar are displaced harmonic-oscillator
wave packets,
\begin{equation}
\eta_n(y,y_0)\propto
H_n\!\left(\frac{y-y_0}{\ell_B}\right)
e^{-(y-y_0)^2/2\ell_B^2},
\end{equation}
whose maxima are centered at the guiding coordinate $y_0$.
The boundary condition does not restrict the spatial extent of these states.
Rather, it imposes a linear constraint at $y=\pm L_y/2$ that selects a discrete
set of admissible guiding centers $\{y_{0,nj}\}$.

A crucial point is that for all such allowed values, the corresponding Landau
wavefunctions retain substantial weight outside the nominal interval
$|y|\le L_y/2$.
As shown in the right-hand panels of Figs.~\ref{yonQuant} and \ref{Robinyon},
approximately half of the probability density lies beyond the geometric sample
edges, independent of whether Dirichlet, Neumann, or Robin constraints are
imposed.
Thus, boundary conditions in the quantum Hall problem do not act as confining
walls; instead, they generate a discrete guiding-center spectrum that underlies
edge-state quantization.

\vspace{0.15cm}
\noindent\textbf{3.\ Absence of barrier repulsion.}
In an ordinary infinite square well, hard walls produce a characteristic
repulsion of eigenvalues, with level spacings that grow with energy.
No analogous effect occurs for Landau states in a magnetic field.
As shown in Figs.~\ref{yonQuant} and \ref{Robinyon}, the guiding-center spectra
selected by Dirichlet, Neumann, and Robin boundary conditions form nearly
uniformly spaced sets, with no systematic repulsion from the edges.

This reflects the fact that the kinetic spectrum remains purely Landau-like,
while the boundary conditions act only as linear constraints selecting
admissible values of the guiding coordinate $y_0$, without introducing an
additional confining potential.
\begin{figure*}[hbt]
\begin{center}
\includegraphics[width=17cm]{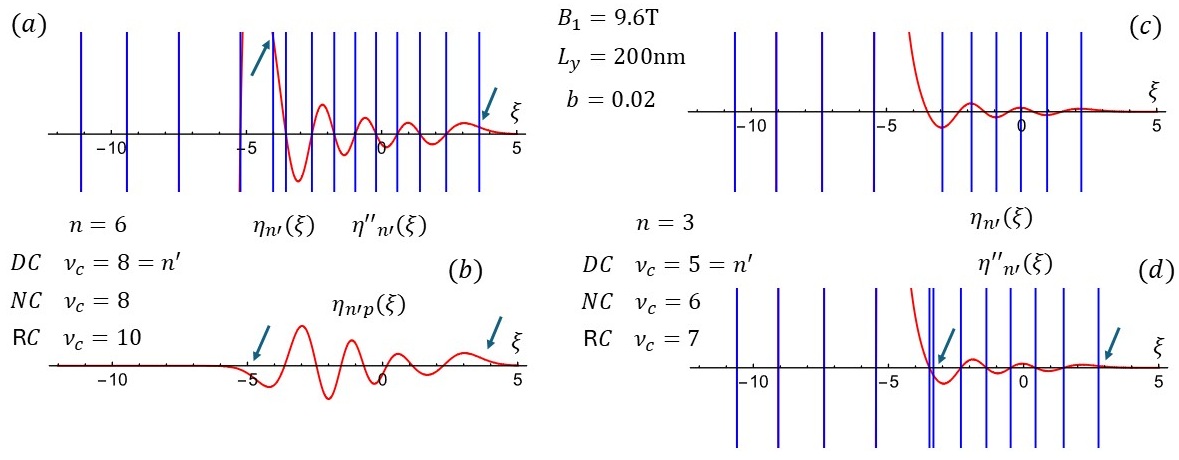}
\caption{
Identification of Dirichlet (DC), Neumann (NC), and Robin (RC) admissible zeros
in the presence of parity breaking.
Shown are exact parity-broken solutions $\eta_{n'}(y)$ of
Eq.~(\ref{ParitySchEq}) together with their third-order perturbative
approximations $\eta_{n'p}(y)$ for $B_1=9.6$~T, $L_y=200$~nm, and $b=0.02$.
Panels (a,b) correspond to $n=6$, while panels (c,d) correspond to $n=3$.
Red curves denote the wave functions, and blue curves their second derivatives.
Arrows mark inflection points where $\eta''_{n'}(y)=0$ with
$\eta_{n'}(y)\neq 0$, which determine the additional Robin-admissible roots.
Panel (c) illustrates Neumann zeros ($\eta'_{n'}=0$), while panel (d) shows
Dirichlet zeros ($\eta_{n'}=0$) together with the outer Robin roots.
This inflection-point criterion provides a robust graphical method for counting
conducting edge channels in the parity-broken case.
}
\label{DNRPBzeros}
\end{center}
\end{figure*}
\vspace{0.15cm}
\noindent\textbf{4.\ Same density, different multiplicity.}
For a fixed Landau index $n$, Dirichlet boundary conditions yield $n$
admissible guiding centers, Neumann conditions yield $n+1$, and the special
Robin condition considered here yields $n+2$.
Despite these differences in multiplicity, Figs.~\ref{yonQuant}(a)--(b) and
Fig.~\ref{Robinyon}(a) show that the \emph{spacing} and \emph{density} of the
allowed $y_{0,nj}$ values are essentially the same in all three cases.

The additional Neumann solution corresponds to a parity-protected root of the
derivative constraint and provides the microscopic origin of an extra edge
branch that survives under weak asymmetry.
The two additional Robin solutions arise near the outer inflection region of the
Gaussian envelope, where the wave packet changes curvature.
These extra boundary-allowed branches enrich the effective edge-channel
counting and generate fractional filling-factor sequences.

\vspace{0.15cm}
\noindent\textbf{5.\ Nearly identical edge wave packets.}
The right-hand panels of Figs.~\ref{yonQuant} and \ref{Robinyon} display
representative edge-localized Landau states (here for $n=4$) evaluated at the
quantized guiding centers obtained from Dirichlet, Neumann, and Robin
conditions.
In all cases the resulting states exhibit the same rapidly oscillating,
edge-localized Gauss--Hermite structure, comparable leakage beyond the nominal
sample boundary, and similar decay lengths.

Thus, the boundary condition does not modify the physical shape of an edge
state; it determines only \emph{how many} such boundary-consistent edge
branches exist.

\vspace{0.15cm}
\noindent\textbf{Summary.}
Dirichlet, Neumann, and Robin boundary conditions should not be interpreted as
hard or soft confining walls for Landau wave functions.
Rather, they serve as microscopic quantization rules for the guiding-center
coordinate of displaced Landau wave packets.
Their primary physical consequence is the boundary-dependent multiplicity of
longitudinal edge-state channels: $n$ for Dirichlet, $n+1$ for Neumann, and
$n+2$ for Robin.

This boundary-controlled channel counting underlies the emergence of both
integer and fractional Hall plateau sequences.
In particular, Neumann and Robin conditions support additional edge branches
that remain robust under weak parity breaking and give rise to the enriched
fractional hierarchy observed experimentally.

\section{Momentum sign, localization, and edge exchange}

Figure~\ref{EdgeStatesMomenta} shows the longitudinal momenta $k_{x,nj}$ of
representative boundary-quantized edge channels as functions of the magnetic
field.
Both the sign and magnitude of $k_{x,nj}$ carry direct physical meaning: for
states localized near the upper boundary, positive $k_x$ corresponds to
propagation along the chiral direction of that edge.

The pinned low-energy channels identified in our analysis are characterized by
large positive $k_x$ and remain tightly localized near the physical boundary as
the magnetic field increases.
These modes therefore persist as robust conduits that sustain the edge current
across successive Landau-level crossings.

In contrast, branches whose longitudinal momentum decreases in magnitude or
changes sign progressively shift away from a given edge and effectively exchange
to the opposite boundary.
Such modes no longer contribute to transport on the original edge, but instead
reappear as counterpropagating channels on the other side of the Hall bar.
\begin{figure*}[hbt]
\begin{center}
\includegraphics[width=16.5cm]{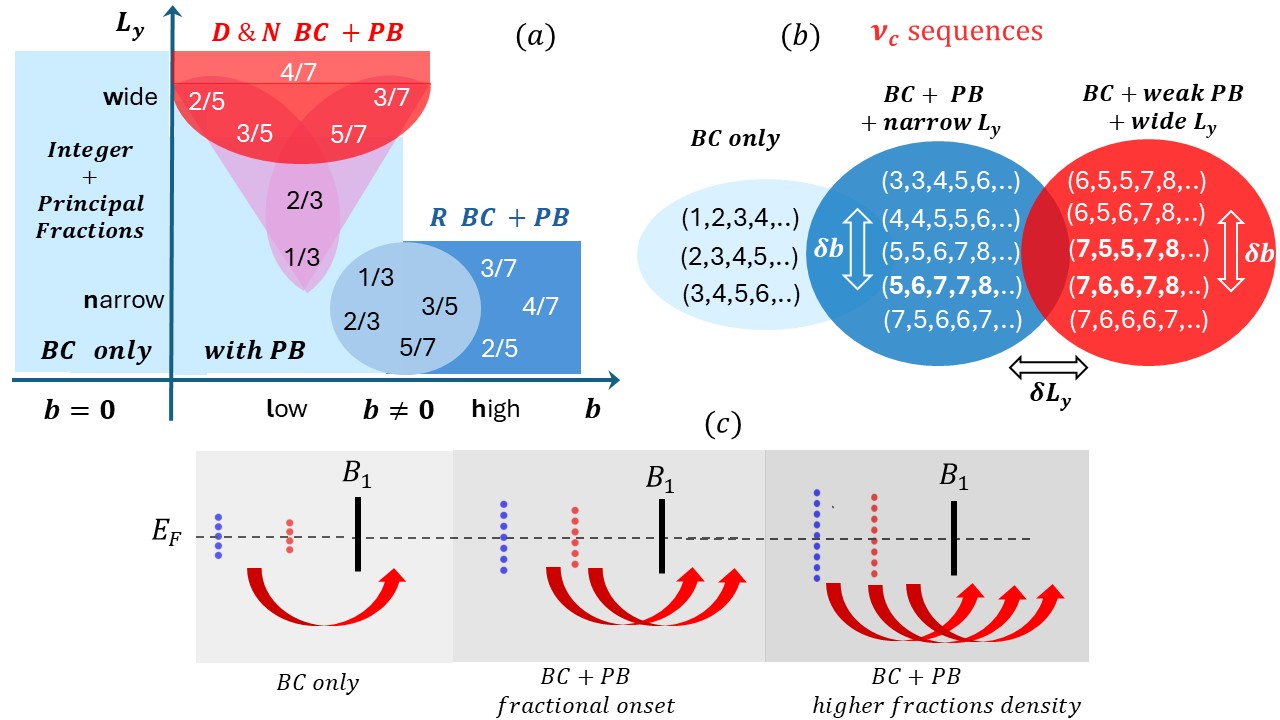}
\caption{Schematic phase space of fractional filling-factor sequences generated by boundary quantization and parity breaking.
(a) Parameter map in the plane of effective lateral width $L_y$ and parity-breaking strength $b$, identifying regimes in which Dirichlet/Neumann boundary conditions (upper region) or Robin boundary conditions (lower region) determine the admissible guiding-center solutions. The light-blue domain corresponds to integer and principal fractional states produced by boundary quantization alone. Boundary quantization sets the baseline multiplicity of edge channels, while parity breaking lifts the degeneracy between opposite edges and reorganizes the available branches. Principal fractions such as $\nu=1/3$ remain stable over broad regions of parameter space, whereas higher-order fractions (e.g., $2/5$, $3/7$, $4/7$) appear only within restricted mixed domains.
(b) Representative sequences of effective channel multiplicities $\nu_c$ in the boundary-condition-dominated, weak-parity-breaking, and wide-sample regimes, illustrating the sensitivity of subleading fractions to variations in $b$ and $L_y$.
(c) Emergence of edge channels pinned near the Fermi level for magnetic fields exceeding the $n=1$ plateau field $B_1$. Parity splitting selectively stabilizes subsets of chiral modes, increasing the density of accessible fractional plateaus at higher fields.}\label{PhaseSpaceDiagram}
\end{center}
\end{figure*}
The momentum evolution in Fig.~\ref{EdgeStatesMomenta} thus provides a
microscopic picture of the \emph{edge-exchange} mechanism.
As the magnetic field varies, the roots of the boundary quantization conditions,
parametrized by the guiding-center coordinates $\xi_{nj}$, shift continuously,
reassigning localization and momentum between the two edges while conserving the
total number of available edge-state channels.

This process preserves the global chiral structure of edge transport and
clarifies why only a subset of parity-split channels remains energetically
accessible and current-carrying in the fractional regime.

\section{Counting zeros in the presence of parity breaking}

As discussed in the main text, the number of conducting edge-state channels
$\nu_c$ plays a central role in determining the effective filling factors that
enter the transport problem.
In the absence of parity breaking, $\nu_c$ is fixed uniquely by the imposed
boundary condition---Dirichlet (DC), Neumann (NC), or Robin (RC).
When a weak parity-breaking interaction is present, however, the number of
admissible edge channels becomes a function of both the asymmetry strength $b$
and the geometric ratio $L_y/\ell_B$.

The exact solutions of the parity-broken Schr\"odinger equation,
Eq.~(\ref{ParitySchEq}), are given in Eq.~(\ref{PBsolution}) in terms of
parabolic cylinder functions.
These generalize the Landau eigenstates and are naturally labeled by a
continuous index $n'$, which provides only an approximate indication of the
number of admissible boundary roots.
A reliable determination of the Dirichlet, Neumann, and Robin zeros therefore
requires a more direct criterion.

In practice, a brute-force numerical root search is cumbersome, since symbolic
solvers return many closely spaced solutions.
Instead, we employ a graphical method based on the structure of the wave packet
and its derivatives, supported by a perturbative representation of the
parity-broken states.

Figure~\ref{DNRPBzeros} compares the exact wave functions $\eta_{n'}(y)$ with
their third-order perturbative approximations $\eta_{n'p}(y)$, together with
their second derivatives.
Although third-order perturbation theory does not reproduce the full amplitude
of the exact solutions, it accurately captures the location of the outer
inflection points, which provide the relevant Robin-admissible roots.

For the Robin condition in the parity-broken case, the additional admissible
zeros are identified by inflection points satisfying
$\eta''_{n'}(y)=0$ while $\eta_{n'}(y)\neq 0$.
For $n=6$ [panels (a,b)], the arrows mark precisely these inflection-point
solutions, which coincide with the outer Robin roots.
For $n=3$, panel (c) illustrates the Neumann zeros defined by $\eta'_{n'}(y)=0$,
while panel (d) shows the Dirichlet zeros together with the corresponding outer
Robin solutions.

Once these inflection points are identified, the number of Dirichlet, Neumann,
and Robin admissible roots---and therefore the number of conducting edge
channels $\nu_c$---can be determined unambiguously in the parity-broken regime.

\section{Phase-space diagram of fractional states}\label{AppendixF}

The diversity of observed fractional quantum Hall plateaus is often described
phenomenologically as a ``zoo'' of fractions. Within the present framework,
however, these states can be organized into a structured hierarchy that reflects
the interplay between boundary-selected guiding-center quantization and weak
parity-breaking edge asymmetry.

Boundary conditions impose a discrete spectrum of admissible guiding centers and
thereby fix the baseline multiplicity of edge-state channels within a given
Landau level. The effective lateral width ($L_y$) controls the density of these
boundary-allowed solutions, while a parity-breaking perturbation lifts the
degeneracy between opposite edges and selectively reorganizes the channel
counting. Together, (($L_y$, $b$)) define a natural parameter space in which
different fractional filling factors correspond to distinct stable edge-channel
configurations, as summarized schematically in Fig.~\ref{PhaseSpaceDiagram}.

Robust principal fractions—most notably ($\nu=1/3$)—arise from top-tier
configurations that persist over extended regions of this parameter space and
are comparatively insensitive to microscopic sample details. By contrast,
higher-order fractions emerge predominantly near the boundaries between
competing admissible multiplicities. In this mixed regime, small variations in
the parity-breaking strength (b) or in the effective width ($L_y$) can enable
or suppress fractions such as (2/5) or (3/7), while leaving the dominant
plateaus essentially unchanged.

This boundary-based organization provides a unified explanation for two
coexisting experimental facts: the remarkable reproducibility of the strongest
fractional states and the pronounced sample dependence of weaker, higher-order
plateaus. In this picture, the observed fractional hierarchy reflects how
interaction-driven correlations are expressed within a guiding-center landscape
that is filtered and structured by boundary quantization and edge asymmetry.


\begin{thebibliography}{100}
\bibitem{vonKlitzing1980} von Klitzing K., Dorda G. \& Pepper M. New Method for High-Accuracy Determination of the Fine-Structure Constant
Based on Quantized Hall Resistance. {\it Phys. Rev. Lett. }  {\bf 45}, 494 (1980)
\bibitem{Thouless1982} Thouless, D. J., Kohmoto, M., Nightingale, M. P., \& den Nijs, M. Quantum Hall Conductance in a Two-Dimensional Periodic Potential. {\it Phys. Rev. Lett.} 49, 405-408 (1982).
\bibitem{Stormer} Tsui, D.C., Störmer, L. \& Gossard, A.C. Two-Dimensional Magnetotransport in the Extreme Quantum Limit {\it Phys. Rev. Lett.} 48, 1559-1562 (1982).

\bibitem{Laughlin1983} Laughlin, R.B. Anomalous Quantum Hall Effect: An Incompressible Quantum Fluid with Fractionally Charged Excitations {\it Phys. Rev. Lett.}  50, 1395-1398 (1983).
\bibitem{Yu2010} Yu, R. , Zhang, W., Zhang, H J., Zhang, S.C., Dai, X. \& Fang, Z. Quantized Anomalous Hall Effect in Magnetic Topological Insulators {\it Science}  329, 61-64
(2010).
\bibitem{McIver2020}. McIver, J.W, Schulte, B., Stein, F.U., Matsuyama, T., Jotzu, G., Meier, G. \& Cavallier A. Light-Induced anomalous Hall effect in graphene. {\it Nat. Phys.} 16, 38-43 (2020).

\bibitem{Hastings2015} Hastings, M. B.\& Michalakis, S. Quantization of Hall Conductance for Interacting Electrons on a Torus. {\it Commun. Math. Phys.} 334, 433-472 (2015).
\bibitem{Buttiker1988} Büttiker  M., Absence of backscattering in the quantum Hall effect in multiprobe conductors {\it Phys. Rev. B }  38, 9375-9389 (1988).
\bibitem{Morse1953} P. M. Morse and H. Feshbach, {\it Methods of Theoretical Physics} (McGraw Hill, 1953), see page 679.

\bibitem{Koch1993} Koch, S., Haug, R.J., von Klitzing, K. \& Razeghi, M. Observation of a spin-polarization instability in tilted magnetic fields. {\it Physica B}  184, 76-80 (1993).
\bibitem{WeisVonKlitzing} Weis, J. \& von Klitzing, K. Metrology and microscopic picture of the integer quantum Hall effect. {\it Phil. Trans. R. Soc. A}  369, 3954-3974 (2011).
\bibitem{Stormer1992} Störmer, H.L. {\it Physica B} 177, 401-408 (1992).
\end{thebibliography}
\end{document}